\documentclass[aps,preprint,prd,nofootinbib,epsfig]{revtex4}

\usepackage{feynman}
\usepackage{pifont}
\usepackage{color}
\usepackage{amssymb}
\usepackage{hyperref}
\usepackage{amsmath}
\usepackage{graphicx}
\usepackage{fancyhdr}
\usepackage[caption=false]{subfig}
\newcommand{\be}{\begin{equation}}
\newcommand{\ee}{\end{equation}}
\newcommand{\bea}{\begin{eqnarray}}
\newcommand{\eea}{\end{eqnarray}}
\newcommand{\bes}{\begin{subequations}}
\newcommand{\ees}{\end{subequations}}

\newcommand{\bc}{\begin{center}}
\newcommand{\ec}{\end{center}}
\newcommand{\bsvar}{\sigma_{b}^2}
\usepackage{multirow}
\usepackage{booktabs}
\usepackage{makecell}
\usepackage{tabulary}
\usepackage{verbatim}
\usepackage{cancel}
\begin{document}
%%%%%
\title{Deep learnig analysis of the  inverse seesaw in a 3-3-1 model at the LHC.}
\author{D. Cogollo$^1$}
\email{diegocogollo@df.ufcg.edu.br}
\author{F. F. Freitas$^{2}$}
\email{felipefreitas@ua.pt}
\author{C. A. de S. Pires$^{3}$}
\email{cpires@fisica.ufpb.br}
\author{Yohan M. Oviedo-Torres$^{1,3}$}
\email{ymot@estudantes.ufpb.br}
\author{P. Vasconcelos$^{4}$}
\email{pablo.wagner@professor.ufcg.edu.br}
\affiliation{$^1$Departamento de F\'isica, Universidade Federal de Campina Grande, Caixa Postal 10071, 58429-900,  Campina  Grande, Para\'\i ba,  Brazil \\
$^2$Departamento de F\'{\i}sica, Universidade de Aveiro and CIDMA, Campus de Santiago, 3810-183 Aveiro, Portugal\\
$^3$Departamento de F\'isica, Universidade Federal da Para\'iba,
Caixa Postal 5008, 58051-970, Jo\~ao Pessoa, PB, Brazil\\
$^4$Unidade Acad\^{e}mica de Engenharia de Produ\c c\~{a}o - CDSA, Universidade Federal de Campina Grande, Caixa Postal 112, 58540-000,
Sum\'{e}, PB, Brazil}
\date{\today}
\begin{abstract}
Inverse seesaw is a genuine  TeV scale seesaw mechanism. In it active neutrinos with masses at eV scale requires lepton number be explicitly violated at keV scale and the existence of new physics, in the form of heavy neutrinos,  at TeV scale. Therefore it is a phenomenologically viable seesaw mechanism since its signature   may be probed at the LHC. Moreover it is  successfully embedded into gauge extensions of the standard model as the  3-3-1 model with the right-handed neutrinos. In this work we revisit the implementation of this mechanism into the 3-3-1 model and employ deep learning analysis to probe such setting at the LHC  and, as main result, we have that if its signature is not detected in the next  LHC running with energy of 14 TeVs, then, the vector boson $Z^{\prime}$ of the 3-3-1 model  must be heavier than 4 TeVs.
\end{abstract}

\maketitle

\section{Introduction}
%%%%%%%%%%%%%%%%%%%%%%%
Seesaw mechanisms\cite{seesawI, seesawII,seesawIII,inverseseesaw} are seem as the simplest proposals to solve the long-standing problem  of the smallness of the neutrino masses. Recently researchers have focused their investigations on phenomenologically viable seesaw mechanisms, as inverse seesaw one\cite{inverseseesaw}, since their signatures may be probed at the LHC\cite{SSLHC}.

The distinguishable aspect of the inverse seesaw (ISS) mechanism is the fact that it is a genuine TeV scale seesaw mechanism and according to the original idea\cite{inverseseesaw} its implementation  requires the addition of six new neutrinos ( $N_{iR} \, ,\, S_{iL}$ with $i=1,2,3$ ) to the standard model particle content  composing the following bilinear  terms\cite{origin},
\begin{equation}
{\cal L}= -\bar \nu_L  m_D N_R  - \bar S_L  M N_R -  \frac{1}{2}\bar S_L  \mu {S_L}^C + H.c.,
\label{massterms}
\end{equation}
where $m_D$, $M$ and $\mu$  are generic $3\times 3 $ complex mass matrices. These terms  can be arranged  in the following  $9\times 9$ neutrino mass matrix  in the basis {\bf $(\nu_L\,,\,N_L^C\,,\,S_L)$},
\begin{equation}
M_\nu=
\begin{pmatrix}
0 & m^T_D & 0 \\
m_D& 0 & M^T\\
0 & M & \mu
\end{pmatrix}.
\label{ISSmatrix}
\end{equation}
Considering the hierarchy  $\mu<< m_D<<M$, the diagonalization of this $9 \times 9$ mass matrix provides  the following effective neutrino mass matrix for the standard neutrinos:
\begin{equation}
m_\nu = m_D^T (M^T)^{-1}\mu M^{-1} m_D.
\label{inverseseesaw}
\end{equation}
The double suppression by the mass scale connected with $M$ turns it possible to have such scale much below than that one involved in the canonical seesaw mechanism\cite{seesawI,seesawII,seesawIII}. It happens that standard neutrinos with mass at sub-eV scale are obtained for $m_D$ at electroweak scale, $M$ at TeV scale and $\mu$ at keV scale. In this case all the new six neutrinos may develop masses around TeV scale or less, and their mixing with the standard neutrinos is modulated by the ratio $m_DM^{-1}$. The core of the ISS mechanism is that the smallness of the neutrino masses is guaranteed by assuming that the $\mu$ scale is small and, in order to bring  heavy neutrino masses down to TeV scale, it has to be at the keV scale.

In this regard it was showed in \cite{ISS331} that the $SU(3)_C \times SU(3)_L \times U(1)_N$ with right-handed neutrinos (331RHN)\cite{original331} has the main ingredients for realizing the ISS mechanism. However, a probe of the ISS mechanism in 331RHN at the LHC is missing. The proposal of this work is to complete this job and probe the ISS in 331RHN at the LHC. For this purpose we review the model, the mechanism, and employ deep learning to probe the signature of the mechanism at the LHC by means of the production of these new neutrinos and their detection in the form of leptons as final products. 

This work is organised as follow: in Sec. II we revised the implementation of the ISS into the 331RHN and  present the charged and neutral currents of interest for our analysis. In Sec. III we perform our analysis by applying deep learning techniques to probe both the ISS and the 331RHN. In Sec. IV we present our conclusions.

%%%%%%%%%%%%%%%%%%%%%%
\section{Some essential points of the model and of the mechanism}
In order to implement the ISS mechanism into the 3311RHN we have to add three left-handed neutral fermions in the singlet form to the original leptonic content of model,
\begin{equation}
{ L }_{ aL }={ \left( \begin{matrix} { \nu  }_{ a } \\ { l }_{ a } \\ { { \nu  } }_{ a }^{ C } \end{matrix} \right)  }_{ L }\sim \left( 1,3,-\frac { 1 }{ 3 }  \right);
\label{ed334}
\end{equation}

\begin{equation}
{ { l } }_{ R }^{ a }\sim \left( 1,1,-1 \right) ,\quad { { N } }_{ L }^{ a }\sim \left( 1,1,0 \right),
\label{ed335}
\end{equation}

where $a=1,2,3$ which corresponds to three families of leptons.

For completeness reasons, we present the quark  content. As it is well known,  in the quark sector, two families must transform as anti-triplet. This is so to cancel anomalies. Here we make the following choice:
\bea
&& Q_{i_L}=
\left (
\begin{array}{c}
d_i \\
-u_i \\
d^{\prime}_i
\end{array}
\right )_L \,\,\,\, \sim (\mbox{{3}},\mbox{{ 3$^*$}},0),\nonumber \\
\newline \nonumber \\
&& u_{i_R} \sim \left(\mbox{{3}},\mbox{{1}},\frac{2}{3}\right)\,\,\, , 
\,\,\,d_{i_R} \sim \left(\mbox{{3}},\mbox{{1}},-\frac{1}{3}\right)\,\,\, , 
\,\,\, d^{\prime}_{i_R} \sim \left(\mbox{{3}},\mbox{{1}},-\frac{1}{3}\right) , 
\label{1.4} 
\eea
where $i=1,2$ while the third family will transfrom as triplet,   
\bea
&&Q_{3_L}=
\left (
\begin{array}{c}
u_3\\
d_3\\
T
\end{array}
\right )_L \,\, \sim \left(\mbox{{3}},\mbox{${3}$},\frac{1}{3}\right),\nonumber \\
\newline \nonumber \\
&& u_{3_R} \sim \left(\mbox{{3}},\mbox{{1}},\frac{2}{3}\right)\,\,\, , 
\,\,\,d_{3_R} \sim \left(\mbox{{ 3}},\mbox{{1}},-\frac{1}{3}\right)\,\,\, , 
\,\,\, T_{R} \sim \left(\mbox{{3}},\mbox{{1}},-\frac{2}{3}\right) \quad.
\label{1.7}
\eea

The scalar sector keeps the original content,
\bea
\eta =
\left (
\begin{array}{c}
\eta^0 \\
\eta^- \\
\eta^{\prime 0}
\end{array}
\right )\sim({1},{3}, -\frac{1}{3}),\,\rho =
\left (
\begin{array}{c}
\rho^+ \\
\rho^0 \\
\rho^{\prime +}
\end{array}
\right ) \sim({1},{3},\frac{2}{3}),\, \chi =
\left (
\begin{array}{c}
\chi^0 \\
\chi^{-} \\
\chi^{\prime 0}
\end{array}
\right ) \sim({1},{3},-\frac{1}{3}).
\nonumber
\eea

The gauge sector is composed by the standard ones , $W^{\pm}_\mu\,,\,Z_\mu$ and the photon $A_\mu$ plus five new ones $U^0_\mu\,,\, U^{0 \dagger}_\mu\,,\,W^{\prime \pm}_\mu$ and $Z^{\prime}_\mu$.

This particle content allows the following  Yukawa interactions,
\begin{eqnarray}
-{\cal L}^Y &=&f_{ij} \bar Q_{i_L}\chi^* d^{\prime}_{j_R} + {(f_{33}\, \bar Q_{3_L}\,\chi \,T_R)} + g_{ia}\bar Q_{i_L}\eta^* d_{a_R} \nonumber \\
&+& h_{3a} \bar Q_{3_L}\eta u_{a_R} +g_{3a}\bar Q_{3_L}\rho d_{a_R}+h_{ia}\bar Q_{i_L}\rho^* u_{a_R}+ y_{a}\bar L_{a_L} \rho e_{a_R} \nonumber \\
&-& \frac{1}{2}G_{ab}\epsilon_{lmn}\overline{\left(L_{aL}\right)_{l}^c}\rho^*_m (L_{bL})_{n}+G^{\prime}_{ab}\bar L_{aL} \chi (N_{bL})^C + \frac{1}{2} \overline{(N_{L})^C} \mu N_{L} +  \mbox{H.c},
\label{yukawa}
\end{eqnarray}
where $a,b=1,2,3$, $i,j=1,2$  and $l,m,n=1,2,3$. For the sake of simplicity, we consider charged leptons in a diagonal basis. Observe that the last line of this lagrangian includes the terms that trigger the ISS mechanism. 

As usual, we assume that only $\eta^0$, $\rho^0$ and $\chi^{\prime 0}$ develop vaccum expectation values (VEVs) other than zero and we  consider the following  expansions around the VEVs:
\begin{eqnarray}
 \eta^0 , \rho^0 , \chi^{\prime 0} \rightarrow  \frac{1}{\sqrt{2}} (v_{\eta ,\rho ,\chi^{\prime}} 
+R_{ \eta ,\rho ,\chi^{\prime}} +iI_{\eta ,\rho ,\chi^{\prime}}).
\label{vacuaII} 
\end{eqnarray}

With this set of VEVs, the last line of the Yukawa Lagrangian above provides the following mass terms for the neutrinos:
\begin{equation}
{\cal L}_{\nu_{mass}}=\bar \nu_R m_D \nu_L+  \bar{\nu}_{R} M N_{L} + \frac{1}{2} \overline{({N}_{L})^{c}} \mu N_{L} + H.c.
\label{yukawa}
\end{equation}

where the $3\times3$ matrices are defined as
\begin{eqnarray}
& & M_{ab}=G^{\prime}_{ab}\frac{v_{\chi^{\prime}}}{\sqrt2}\label{mM}\\
& & m_{Dab}=G_{ab}\frac{v_\rho}{\sqrt{2}}
\label{mmatrix3x3}
\end{eqnarray}

with $M_{ab}$ and $m_{Dab}$ being Dirac mass matrices, with this last one being antisymmetric.

Considering the basis  ${ S }_{ L }=\left( { \nu  }_{ L },{ \left( { \nu  }^{ C } \right)  }_{ L },{ { N }  }_{ L } \right)$ we can write ${\cal L}_{\nu_{mass}}$ in the following form
\begin{equation}
{\cal L}_{\nu_{mass}}=\frac{1}{2} \overline{(S_L)^c} M_\nu  S_L + H.c.,
\label{massterm}
\end{equation}
with the mass matrix $M_\nu$ having the texture,
\begin{equation}
M_\nu=
\begin{pmatrix}
0 & m^T_D & 0 \\
m_D& 0 & M^T\\
0 & M & \mu
\end{pmatrix}.
\label{massmatrix331}
\end{equation}
This is the mass matrix that characterize the ISS mechanism. The hierarchy  $M \gg
m_D \gg \mu $ provides a seesaw relation for the masses of the standard neutrinos.  In order to see this it is useful to define the matrices,
\begin{equation}
{\cal M}_{D_{6\times3}}=
\begin{pmatrix}
m_{D_{3\times3}} \\
0_{3\times3}
\end{pmatrix},\,\,\,\,\, {\cal M}_{R_{6\times6}}=
\begin{pmatrix}
0_{3\times3} & M^T_{3\times3}  \\
M_{3\times3} &  \mu_{3\times3}
\end{pmatrix},
\label{definition}
\end{equation}
so that we have the following block matrix  where ${\cal M}_R$ is supposed invertible,
\begin{equation}
M_{\nu_{9\times9}}=
\begin{pmatrix}
0_{3\times3} & {\cal M}^T_{D_{3\times6}}  \\
{\cal M}_{D_{6\times3}} & {\cal M}_{R_{6\times6}}
\end{pmatrix}
\label{redefinedmassmatrix331}
\end{equation}
This last matrix can be block diagonalized. For this purpose let us definife the matrix $W$, 

\begin{equation}
W\simeq \left( \begin{matrix} 1-\frac { 1 }{ 2 } { \left( { \mathcal{M} }_{ D } \right)  }^{ \dagger  }{ \left[ { \mathcal{M} }_{ R }{ \left( { \mathcal{M} }_{ R } \right)  }^{ \dagger  } \right]  }^{ -1 }{ \mathcal{M} }_{ D } & { \left( { \mathcal{M} }_{ D } \right)  }^{ \dagger  }{ \left[ { \left( { \mathcal{M} }_{ R } \right)  }^{ \dagger  } \right]  }^{ -1 } \\ -{ \left( { \mathcal{M} }_{ R } \right)  }^{ -1 }{ \mathcal{M} }_{ D } & 1-\frac { 1 }{ 2 } { \left( { \mathcal{M} }_{ R } \right)  }^{ -1 }{ \mathcal{M} }_{ D }{ \left( { \mathcal{M} }_{ D } \right)  }^{ \dagger  }{ \left[ { \left( { \mathcal{M} }_{ R } \right)  }^{ \dagger  } \right]  }^{ -1 } \end{matrix} \right) 
\label{mixW}
\end{equation}

such that,

\begin{equation}
W^TM_{\nu_{9\times9}} W=
\begin{pmatrix}
m_{light_{_{3\times3}}} & 0_{3\times6}  \\
0_{6\times3} & m_{heavy_{_{6\times6}}}
\end{pmatrix},
\label{dmatrix}
\end{equation}
where $m_{light}=-{\cal M}_D^T {\cal M}^{-1}_R {\cal M}_D$ and $m_{heavy}={\cal M}_R$. When we plug ${\cal M}_D$ and ${\cal M}^{-1}_R$ in $m_{light}$ we obtain the canonical inverse seesaw mass expression for the standard neutrinos:
\be
m_{light}=m_D^T (M)^{-1}\mu (M^T)^{-1} m_D
\ee

Observe that the matrix in Eq. \eqref{dmatrix} is not diagonal. It is a block diagonal matrix. The  diagonalization of the mass matrix in Eq.  \eqref{redefinedmassmatrix331} is done through the unitary matrix  $V=WU$, such that $V^{T} M_{\nu_{9\times9}} V=m_{diag}$, with $U$ defined as:

\be 
U=\left(\begin{array}{cc}
U_{PMNS} & 0\\ 
 0 &  U_R
      \end{array}\right),
      \label{matrixfinal2}
\ee
with $U_{PMNS}$ being  the PNMS matrix that diagonalizes $m_{light}$ while $U_R$  diagonalizes $m_{heavy}$, and $m_{diag}$ is the diagonal mass  matrix with nine eigenvalues. 

The explicit form of  $V$ is 

\begin{equation} 
V\simeq \left( \begin{matrix} \left[1-\frac { 1 }{ 2 } { \left( { \mathcal{M} }_{ D } \right)  }^{ \dagger  }{ \left[ { \mathcal{M} }_{ R }{ \left( { \mathcal{M} }_{ R } \right)  }^{ \dagger  } \right]  }^{ -1 }{ \mathcal{M} }_{ D }\right]U_{PNMS} & { \left( { \mathcal{M} }_{ D } \right)  }^{ \dagger  }{ \left[ { \left( { \mathcal{M} }_{ R } \right)  }^{ \dagger  } \right]  }^{ -1 }U_{R} \\ -{ \left( { \mathcal{M} }_{ R } \right)  }^{ -1 }{ \mathcal{M} }_{ D }U_{PNMS} & \left [1-\frac { 1 }{ 2 } { \left( { \mathcal{M} }_{ R } \right)  }^{ -1 }{ \mathcal{M} }_{ D }{ \left( { \mathcal{M} }_{ D } \right)  }^{ \dagger  }{ \left[ { \left( { \mathcal{M} }_{ R } \right)  }^{ \dagger  } \right]  }^{ -1 }\right]U_{R} \end{matrix} \right).
\label{mix1}
\end{equation}

In the end of the day we have
\be  
U^T W^T M_\nu WU=\left(\begin{array}{cc}
m_\nu & 0\\ 
 0 &  m_R
      \end{array}\right),
      \label{matrixfinal}
\ee

with $m_\nu=diag(m_1\,,\,m_2\,,\,m_3)$ and $m_R=diag(m_4\,,\,....\,,\,m_9)$. 

The matrix $V$ connects the flavor basis ${ S }_{ L }=\left( { \nu  }_{ L },{ \left( { \nu  }^{ C } \right)  }_{ L },{  { N }   }_{ L } \right)^T$=$\left( { \nu  }_{ L },{ { \zeta  }  }_{ L } \right)^T$ with the physical one which we call $n_L=(n^0_{iL} \,,\,n^1_{kL})^T$ where $n^0_{i_L}$ with $i=1,2,3$ and $n^1_{k_L}$ with $k=1,2,...,6$. The relation between flavor and mass eigenstates, $S_L=Vn_L$, is given explicitly by.

\begin{align}
{ \nu  }_{ aL }&={ \left\{ { U }_{ PMNS }-\frac { 1 }{ 2 } { \left( { \mathcal{M} }_{ D } \right)  }^{ \dagger  }{ \left[ { \mathcal{M} }_{ R }{ \left( { \mathcal{M} }_{ R } \right)  }^{ \dagger  } \right]  }^{ -1 }{ \mathcal{M} }_{ D }{ U }_{ PMNS } \right\}  }_{ ai } { n }_{ iL }^{ 0 } \nonumber \\
&+{ \left\{ { \left( { \mathcal{M} }_{ D } \right)  }^{ \dagger  }{ \left[ { \left( { \mathcal{M} }_{ R } \right)  }^{ \dagger  } \right]  }^{ -1 } { U }_{ R } \right\}  }_{ ak }{ n }_{ kL }^{ 1 };
\label{flavor-mass}
\end{align}

\begin{align}
{ \zeta  }_{ bL }&={ \left\{ \left[ -{ \left( { \mathcal{M} }_{ R } \right)  }^{ -1 }{ \mathcal{M} }_{ D } \right] { U }_{ PMNS } \right\}  }_{ bi }{ n }_{ iL }^{ 0 } \nonumber \\
&+{ \left\{ { U }_{ R }-\frac { 1 }{ 2 } { \left( { \mathcal{M} }_{ R } \right)  }^{ -1 }{ \mathcal{M} }_{ D }{ \left( { \mathcal{M} }_{ D } \right)  }^{ \dagger  }{ \left[ { \left( { \mathcal{M} }_{ R } \right)  }^{ \dagger  } \right]  }^{ -1 }{ U }_{ R }  \right\}  }_{ bk }{ n }_{ kL }^{ 1 }.
\label{flavor-mass1}
\end{align}

For simplicity, we will define the matrix $V$ in the following  form:

\begin{equation}
V=\begin{pmatrix} { V }^{ \nu \nu  } & { V }^{ \nu N } \\ { V }^{ N\nu  } & { V }^{ NN } \end{pmatrix}.
\label{Vmatrix}
\end{equation}
%%%%%%%%%%%%%%%

Returning to $m_{light}$,  on substituting  $m_D=\frac{G}{\sqrt{2}}v_\rho$ and $M=\frac{G^{\prime}}{\sqrt{2}}v_{\chi^{\prime}}$, we obtain
\begin{equation}
m_{light} =\left( G^T (G^{ \prime T})^{-1}\mu (G^{ \prime})^{ -1} G\right)\frac{v^2_\rho }{v^2_{\chi^{\prime}}}.
\label{inverseseesaw331}
\end{equation}

Remember that  $G$ is an anti-symmetric matrix implying that one eigenvalue of the neutrino mass matrix in Eq.~(\ref{inverseseesaw331}) is null.

Solar, reactor, accelerator and atmospheric neutrino experiments have determined\cite{neutrinodata},
\begin{eqnarray}
&& \Delta m^2_{21}\simeq 7.59  \times10^{-5}\mbox{ eV}^2\,,\,  \Delta m^2_{31}\simeq  2.43 \times 10^{-3}\mbox{eV}^2,\nonumber \\
&& \sin^2(2\theta_{12})\simeq 0.86 \,\,\,,\,\,\, \sin^2(2\theta_{23})\simeq 0.92\,\,\,,\,\,\,\sin^2(2\theta_{13})\simeq  0.092. 
\label{currentneutrinodata}
\end{eqnarray}
Moreover, the current status of neutrino physics allows that  at least one of the three neutrinos may be massless. 

Returning to our model, in it the masses of the active neutrinos are obtained by diagonalizing $m_{light}$  in Eq.~(\ref{inverseseesaw331}) which involves many free parameters in the form of Yukawa couplings $G$ and $ G^{\prime}$.
With such a large set of free parameters, there is a great deal of possible solutions that lead to the correct neutrino mass spectrum and mixing in Eq. (\ref{currentneutrinodata}). However due to  the non-unitarity of the mixing matrix $V^{\nu\nu}$ any  set of values for the entries in $G$ and $G^{\prime}$ that do the job must obey the following constraints \cite{nonunitarity},
\begin{equation}
|\eta|<
\begin{pmatrix}
2.0\times 10^{-3} & 3.5\times 10^{-5}  & 8.0\times 10^{-3}  \\
3.5\times 10^{-5} & 8.0\times 10^{-4}  & 5.1\times 10^{-3} \\
8.0\times 10^{-3}  & 5.1\times 10^{-3}  & 2.7\times 10^{-3}
\end{pmatrix},
\label{nonunitatitybounds}
\end{equation}

where $\eta=\frac { 1 }{ 2 } { \left( { \mathcal{M} }_{ D } \right)  }^{ \dagger  }{ \left[ { \mathcal{M} }_{ R }{ \left( { \mathcal{M} }_{ R } \right)  }^{ \dagger  } \right]  }^{ -1 }{ \mathcal{M} }_{ D }$ .

To simplify our job  we consider  $v_\eta=v_\rho=v$. Thus, the constraint   $v^2_\eta + v^2_\rho=(246 \mbox{GeV})^2$ implies  $v=174$GeV. It is supposed that  $v_{\chi^{\prime}}$ lies around TeV. Here we assume   $5$~TeV. We also  consider   ${\bf \mu=0.3\,{\bf{I}}}$ keV where ${\bf{I}}$ is the identity matrix.

Regarding the Yukawa couplings  $G$ and $G^{\prime}$, we consider the scenario   where $G^{\prime}$ is diagonal but non-degenerate and as illustrative case we take,

\begin{equation}
\begin{scriptsize}
G'=\left( \begin{matrix} { g }'_{ 11 } & 0 & 0 \\ 0 & { g }'_{ 22 } & 0 \\ 0 & 0 & { g' }_{ 33 } \end{matrix} \right) \simeq \left( \begin{matrix} 0.019 & 0 & 0 \\ 0 & 0.07 & 0 \\ 0 & 0 & 0.04 \end{matrix} \right),
\label{ed411}
\end{scriptsize}
\end{equation}
and 
\begin{equation}
\begin{scriptsize}
G=\left( \begin{matrix} 0 & { g }_{ 12 } & { g }_{ 13 } \\ { -g }_{ 12 } & 0 & { g }_{ 23 } \\ { -g }_{ 13 } & { -g }_{ 23 } & 0 \end{matrix} \right) \simeq \left( \begin{matrix} 0 & 4.26\times 10^{-3} & 4.97\times 10^{-3} \\ -4.26\times 10^{-3} & 0 & 6.62\times 10^{-3} \\ -4.97\times 10 ^{-3} & -6.62\times 10^{-3} & 0 \end{matrix} \right),
\label{ed412}
\end{scriptsize}
\end{equation}

With these set of values for  $G$, $G^{\prime}$ and for the values of the  VEVs $v$, $v_{\chi^{\prime}}$ and $\mu$ presented above, the diagonalization of the mass matrix $m_{light}$ in Eq. (\ref{inverseseesaw331}) furnishes 
\begin{equation}
m_1 \simeq 0, \quad\quad m_2 \approx  8.7 \times 10^{-3}\mbox{eV},\quad\quad m_3 \approx 4.8\times 10^{-2}\mbox{eV}, 
\label{eigenvalues1}
\end{equation}
with
\begin{equation}
U_{PMNS}\simeq 
\begin{pmatrix}
0.80 & 0.58  & 0.12  \\
-0.48 & 0.52  & 0.70 \\
0.34  & -0.62  & 0.70
\end{pmatrix}.
\label{PMNSprediction}
\end{equation}
This $U_{PMNS}$ implies in the following mixing angles $\theta_{12} = 36^o$, $\theta_{23} =
 45^o$ and   $\theta_{13} = 7^o$ which recover the experimental values in Eq. (\ref{currentneutrinodata}). 
 
Let us check if the values for $G$ and $G^{\prime}$ above are in accordance with non-unitarity constraint\cite{nonunitarity}.

On substituting the set of values of  $G$ and $G^{\prime}$ in $\eta$ yields,

\begin{equation}
\eta =\left( \begin{matrix} 9.6\times { 10 }^{ -6 } & 1.0\times { 10 }^{ -5 } & 3.0\times { 10 }^{ -6 } \\ 1.0\times { 10 }^{ -5 } & 4.3\times { 10 }^{ -5 } & 3.4\times { 10 }^{ -5 } \\ 3.0\times { 10 }^{ -6 } & 3.4\times { 10 }^{ -5 } & 4.5\times { 10 }^{ -5 } \end{matrix} \right) ,
\label{nonunitatityprediction}
\end{equation}
which  respect the bounds in Eq.~(\ref{nonunitatitybounds}). 

 Regarding the six new neutrinos, by diagonalizing $m_{heavy}={\cal M}_{R}$ in Eq.~(\ref{definition}),  our illustrative example yields  ($n^1_{1_L}\,,\,n^1_{6_L},$) with masses $\sim$ 373.28~GeV,  ($n^1_{2_L}\,,\, n^1_{5_L}$) with masses  $\sim$ 220.84~GeV and  ($n^1_{3_L}\,,\,n^1_{4_L}$) with masses around  $\sim$ 96.32~GeV. The degeneracy in mass is due to the simplicity of our illustraive example.

%Morevover, the mixing matrix, $U_R$, that provides the mixing among these neutrinos has the pattern,

%\begin{equation}
%{ U }_{ R }=\left( \begin{matrix} 0 & 0 & 0.7071 & 0.7071 & 0 & 0 \\ -0.7071 & 0 & 0 & 0 & 0 & -0.7071 \\ 0 & 0.7071 & 0 & 0 & 0.7071 & 0 \\ 0 & 0 & -0.7071 & 0.7071 & 0 & 0 \\ 0.7071 & 0 & 0 & 0 & 0 & -0.7071 \\ 0 & -0.7071 & 0 & 0 & 0.7071 & 0 %\end{matrix} \right),
%\label{URprediction}
%\end{equation}

So we developed the basic aspects  of the implementation of the ISS mechanism within the 331RHN and presented an illustrative example that recovers the current experimental results involving neutrino oscillation. 

Our wish now is to probe this scenario at the LHC. We do this by means of the production of pairs of heavy neutrinos, $n^1_{i_L}$,  and their subsequent detection in the form of leptons as main final products. The processes we study are intermediated by the standard charged gauge boson $W^{\pm}$ and  $Z^{\prime}$. The   neutral and charged currents  of interest are presented below.

We present, first, the charged current  with $W^{\pm}$ which are composed by the following terms, 

\begin{equation}
\begin{split}
\mathcal{L}_{n \ell W} = &  -\frac { g }{ \sqrt { 2 }  } \sum _{ a=1 }^{ 3 } \sum _{ i=1 }^{ 3 } \bar { \ell  } _{ aL }\gamma ^{ \mu  }\left[ { U }_{ PMNS }-\frac { 1 }{ 2 } { \left( { { \mathcal{M} } }_{ D } \right)  }^{ \dagger  }{ \left[ { \mathcal{M} }_{ R }{ \left( { \mathcal{M} }_{ R } \right)  }^{ \dagger  } \right]  }^{ -1 }{ \mathcal{M} }_{ D }{ U }_{ PMNS } \right] _{ai}n^{0}_{iL}W_{\mu}^-\\
&- \dfrac{g}{\sqrt{2}}\sum_{a=1}^{3}\sum_{k=1}^{6}\bar{\ell}_{aL}\gamma^{\mu}\left[  { \left( { \mathcal{M} }_{ D } \right)  }^{ \dagger  }{ \left[ { \left( { \mathcal{M} }_{ R } \right)  }^{ \dagger  } \right]  }^{ -1 } { U }_{ R }\right] _{ ak }n^{1}_{kL}W_{\mu}^- +H.c. 
\label{CC}
\end{split}
\end{equation}

%Now we present the charged currents with the new charged gauge boson $W^{\prime +}$. It is not the dominant one. We present it here for sake of completeness
%\bea
%{\cal L}^{W^{\prime}} &=& -\frac{g}{\sqrt{2}}\bar e_{a_L} \gamma^\mu [(1-\frac{1}{2}F^{\dagger}F)U_R ]_{ak}n^1_{k_L} W^{\prime +}_\mu+\nonumber \\
%&&-\frac{g}{\sqrt{2}}\bar e_{a_L} \gamma^\mu [ ({\cal M}^{-1}_R)^T {\cal M}_D U_{PMNS}]_{ai} n^0_{i_L} W^{\prime +}_\mu +H.c
%\eea
%The neutral currents with the standard gauge boson $Z$ are these ones {\bf acho que podemos apagar as interaçoes do Z, nao sao usadas na analise}:
%\begin{equation}
%\begin{split}
%\mathcal{L}_{\nu\nu Z} = &- \dfrac{g}{2\cos{\theta_W}} [\sum_{i,j=1}^{3} \bar{n^0}_{iL} (V^{\nu\nu\dagger}V^{\nu\nu})_{ij}\gamma^{\mu}n^0_{jL} + \sum_{i=1}^{3}\sum_{m=1}^{6}\bar{n^0}_{iL} (V^{\nu\nu\dagger}V^{\nu N})_{im}\gamma^{\mu}n^{1}_{mL}\\
%& + \sum_{k=1}^{6}\sum_{j=1}^{3}\overline{n^1}_{kL}(V^{\nu N\dagger}V^{\nu\nu})_{kj}\gamma^{\mu}n^0_{jL} + \sum_{k=1}^{6}\sum_{m=1}^{6}\overline{n^1}_{jL}(V^{\nu N\dagger}V^{\nu N})_{km}\gamma^{\mu}n^1_{mL}]Z_{\mu}.
%\end{split}
%\label{znunu}
%\end{equation}

The neutral current interactions with $Z^{\prime}$ have two contributions. The first one is
\begin{equation}
\begin{split}
\mathcal{L}_{nnZ^{\prime}} = &- \mathcal{G}\dfrac{g}{2\cos{\theta_W}}[\sum_{i,j=1}^{3} \bar{n^0}_{iL} (V^{\nu\nu\dagger}V^{\nu\nu})_{ij}\gamma^{\mu}n^0_{jL} + \sum_{i=1}^{3}\sum_{m=1}^{6}\bar{n^0}_{iL} (V^{\nu\nu\dagger}V^{\nu N})_{im}\gamma^{\mu}n^{1}_{mL}\\
& + \sum_{k=1}^{6}\sum_{j=1}^{3}\overline{n^1}_{kL}(V^{\nu N\dagger}V^{\nu\nu})_{kj}\gamma^{\mu}n^0_{jL} + \sum_{k=1}^{6}\sum_{m=1}^{6}\overline{n^1}_{kL}(V^{\nu N\dagger}V^{\nu N})_{km}\gamma^{\mu}n^1_{mL}]Z^{\prime}_{\mu},
\end{split}
\label{zprimenunu}
\end{equation}

with $\mathcal{G}=\frac{1-2\sin^2\theta_{ W }}{\sqrt {3-4\sin^2\theta_{W}}}$ and,

\begin{equation}
\begin{split}
\mathcal{L}_{nnZ^{\prime}} = & \mathcal{F}\dfrac{g}{2\cos{\theta_W}}[\sum_{i,j,b=1}^{3} \bar{n^0}_{iL} (V^{N\nu})_{bi}^{\star}(V^{N\nu})_{bj}\gamma^{\mu}n^0_{jL} + \sum_{i,b=1}^{3}\sum_{m=1}^{6}\bar{n^0}_{iL} (V^{N\nu})_{bi}^{\star}(V^{NN})_{bm}\gamma^{\mu}n^{1}_{mL}\\
& + \sum_{k=1}^{6}\sum_{b,j=1}^{3}\overline{n^1}_{kL}\gamma^{\mu}(V^{NN})_{bk}^{\star}(V^{N\nu})_{bj}n^0_{jL} + \sum_{k,m=1}^{6}\sum_{b=1}^{3}\overline{n^1}_{kL}(V^{NN})_{bk}^{\star}(V^{NN})_{bm}\gamma^{\mu}n^1_{mL}]Z^{\prime}_{\mu},
\end{split}
\label{zprimenunu1}
\end{equation}

with $\mathcal{F}=\frac{2\cos^2\theta_{ W }}{ \sqrt {3-4\sin^2\theta_{W}}}$.

 This is the set of interactions that matter for us here. In the first line of Eq. (\ref{CC}) we have the mixing matrix $\left ( 1-\frac { 1 }{ 2 } { \left( { \mathcal{M} }_{ D } \right)  }^{ \dagger  }{ \left[ { \mathcal{M} }_{ R }{ \left( { \mathcal{M} }_{ R } \right)  }^{ \dagger  } \right]  }^{ -1 }{ \mathcal{M} }_{ D }\right ){ U }_{ PMNS } =V^{\nu \nu}$.  Due to the smallness of the second term, see values in Eq.  \eqref{nonunitatityprediction}, we take $V^{\nu \nu} \simeq U_{PMNS}$.

In the second line of Eq. (\ref{CC}) there appear the mixing matrix $({ \left( { \mathcal{M} }_{ D } \right)  }^{ \dagger  }{ \left[ { \left( { \mathcal{M} }_{ R } \right)  }^{ \dagger  } \right]  }^{ -1 } { U }_{ R })_{  }=V^{\nu N}$. Our illustrative example yields,
%%%%%
\begin{equation}
{ V }^{ \nu N  }\simeq \left( \begin{matrix} -1.4 \times 10^{-3}\,\,\,\,\,\,\,\,\,\,\, & 2.8 \times 10^{-3} & 0 & 0 & -2.8 \times 10^{-3} \,\,\,\,\,\,\,\,\,\,\,& 1.4 \times 10^{-3} \\ 0 & 3.7 \times 10^{-3} & -5.5 \times 10^{-3} & 5.5 \times 10^{-3} & -3.7 \times 10^{-3} & 0 \\ 2.2\times 10^{-3} & 0 & -6.3 \times 10^{-3} & 6.3 \times 10^{-3} & 0 & -2.2\times 10^{-3} \end{matrix} \right);
\label{ed422}
\end{equation}
%%%%
Such pattern of mixing is due to the simple choice of the parameters  $G^{\prime}$ and $\mu$. In the next section we are going to probe the signature of this mechanism  by producing  the lightest new neutrinos, $n^1_{3_L}$ and $n^1_{4_L}$, at the LHC. Observe that as $(V^{\nu N})_{1 3}$ and $(V^{\nu N})_{1 4}$ are null, then these neutrinos do not form charged currents with the electrons. For this reason the analisys done in the next section is based on the production of these neutrinos and their final products in the form of muons .

 Concerning  neutral currents, we also explore the direct production of $Z^{\prime}$ and its subsequent decay into a pair of $n^1_{3_L}$ or $n^1_{4_L}$. The interactions that generate these processes are the last terms of the Eqs. \eqref{zprimenunu} and \eqref{zprimenunu1}. Our illustrative example yields the following values for the mixing matrix $V^{N N}$,

\begin{equation}
\begin{scriptsize}
V^{N N} \simeq \left( \begin{matrix} 0 & 0 & 7.0 \times 10^{-1} & 7.0 \times 10^{-1} & 0 & 0 \\ -7.0 \times 10^{-1} & 0 & 0 & 0 & 0 & -7.0 \times 10^{-1} \\ 0 & 7.0 \times 10^{-1} & 0 & 0 & 7.0 \times 10^{-1} & 0 \\ -3.91\times { 10 }^{ -5 } & -5.68\times { 10 }^{ -5 } & -7.0 \times 10^{-1} & 7.0 \times 10^{-1} & 5.68\times { 10 }^{ -5 } & 3.91\times { 10 }^{ -5 } \\ 7.0 \times 10^{-1} & 1.10\times { 10 }^{ -5 } & 3.91\times { 10 }^{ -5 } & -3.91\times { 10 }^{ -5 } & -1.10\times { 10 }^{ -5 } & -7.0 \times 10^{-1} \\ -1.10\times { 10 }^{ -5 } & -7.0 \times 10^{-1} & -5.68\times { 10 }^{ -5 } & 5.68\times { 10 }^{ -5 } & 7.0 \times 10^{-1} & 1.10\times { 10 }^{ -5 } \end{matrix} \right),
\end{scriptsize}
\label{ed424}
\end{equation}
that along with Eq. \eqref{ed422} allows us to perform the analysis for this production. 

Before go into the analysis, with the charged and neutral currents at hand, first thing to do is to  check  if our illustrative example obeys the rare   lepton flavor violation(LFV) process $\mu \rightarrow e\gamma$ constraint. Such process is allowed by the second coupling in Eq.~(\ref{CC}). The branching ratio for the process mediated by these  six heavy neutrinos is given by\cite{LFV},
\begin{equation}
BR(\mu \rightarrow e \gamma)\approx \frac{\alpha^3_W \sin^2(\theta_W) m^5_{\mu}}{256 \pi^2 m^4_W \Gamma_\mu}\times |\sum_{k=1}^6 {(V^{\nu N})_{e k}(V^{\nu N})_{\mu k}I(\frac{m^2_{n_{k L}^1}}{m_{W}^{2}})}|^2,
\nonumber
\end{equation}

where
\begin{equation}
I(x)=-\frac{2x^3+5x^2-x}{4(1-x)^3}-\frac{3x^3\ln x}{2(1-x)^4}.
\end{equation}

In the above branching ratio expression  we use  $\alpha_W= \frac{g^2}{4 \pi}= 3.3 \times 10^{-2}$,  $\sin^2(\theta_W)=0.231$, $m_\mu=105$ ~Mev, $m_W=80.385$~Gev , $\Gamma_\mu=3 \times 10^{-16}$ Mev. The present values of these parameters are found in \cite{PDG}.  Our illustrative example provides $BR(\mu \rightarrow e \gamma)\approx 1.4 \times 10^{-13}$. This is very close to the current bound that is $BR(\mu \rightarrow e \gamma)< 4.2 \times 10^{-13}$\cite{MEG}. So, this case may be confirmed or excluded at the next running of the MEG experiment.

\section{Analysis of the production mechanism and main channels}
\label{production}

There are two major production channels for the $n^1_{i_L}$ neutrinos. The first one is via vector gauge boson $W^{\pm}$, which can be produced trough the s-channel in a proton-proton collision. In the particular case of our illustrative example, the $W^{\pm}$ can further decay into a $\mu$ lepton and the neutrinos $n^1_{i_L}$. On the other hand, the $n^1_{i_L}$ can decay into $\mu$ and $W^{\pm}$. Then this channel can have as final product 3 leptons plus missing energy ($\mu^{\pm}\mu^{\mp}\ell^{\pm}\nu_{\ell}$) or  2 muons and 2 jets ($\mu^{\pm}\mu^{\mp}jj$). 

The second production mechanism for the neutrinos $n^1_{i_L}$ is through the direct production of the $Z^{\prime}$ and its subsequent decay into a pair of $n^1_{i_L}$. The final state for this type of channel will appear as pair of high boosted muons, pair of leptons and missing transverse energy ($\mu^{\pm}\mu^{\mp}\ell^{\mp}\nu_{\ell}\ell^{\pm}\nu_{\ell}$) or pair of high boosted muons and 4 light jets. We investigate both channels and explore the phenomenological features of this model and how the signatures of the $n^1_{i_L}$ can appear at the listed final states at the LHC.

To do so, we generate an UFO \cite{Degrande:2011ua} file using the FeynRules \cite{Alloul:2013bka}. This UFO file is latter used by the MadGraph5 \cite{Alwall:2014hca} package to produce the hard scattering processes we want to investigate. All the hard scattering processes are further pass to Pythia version 8.1 \cite{Sjostrand:2007gs} and Delphes \cite{deFavereau:2013fsa} in order to hadronize and include the detector effects to make the data from of Monte-Carlo pseudo-events be as close as possible to the data produced by the LHC at 14 TeV.

\subsection{$p p \rightarrow \mu^{\pm}\mu^{\mp}e^{\pm}\nu_{e}$ channel:}

\begin{figure}[h!]
\begin{center}
\subfloat{
\includegraphics[clip,scale=0.6]{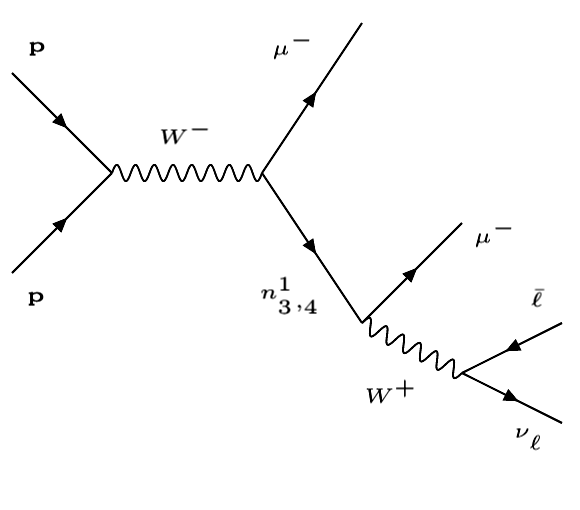}
}
\caption{ Production of $n^1_{(3,4)}$ at the LHC via $W$ channel.}
\label{Wchanell}
\end{center}
\end{figure}

As mentioned earlier, this is one of the main production mechanisms for the production of $n^1_{i_L}$ and is displayed in FIG. (\ref{Wchanell}). To investigate this channel we generate 450000 events with 14 TeV center of mass energy. To stay safely away from infrared and colinear divergences, we apply the basic cuts of Eq.~\eqref{eq:basic_cuts} at the generation level
\begin{eqnarray}
&& p^{\ell}_{T} > 20\; \hbox{GeV},\; p^{j,b}_{T} > 30\; \hbox{GeV},\nonumber\\  
&& \vert\eta^{j,b}\vert < 3.0,\; \vert\eta^{\ell}\vert < 2.7,\nonumber \\
&& \Delta R_{jj,bb,\ell\ell} > 0.01\; .
\label{eq:basic_cuts}
\end{eqnarray}
%%%%

%%%%

We focus our investigation in the production of the lightest new neutrinos. Thus, we are going to analyze  the channel $$pp\rightarrow W^{\pm}\rightarrow \mu^{\pm} n^1_{3_L}(n^1_{4_L}),$$ with the decay chain for the neutrino $$n^1_{3_L}(n^1_{4_L}) \rightarrow \mu^{\pm}W^{\mp}, W^{\pm}\rightarrow e^{\pm}\nu_{e}.$$
This choice allow us to reconstruct, with a good accuracy, the full decay chain generated by the $n^1_{i_L}$. Another reason for this choice stems from the fact that in our model the couplings between $W^{\pm},n^1_{3_L}$ or $W^{\pm},n^1_{4_L}$ and $\mu$ are relatively large, allowing a sizable cross section for the production at the LHC. As consequence for this choice we have as main irreducible background the channels:

\begin{eqnarray}
ZW^{\pm}\rightarrow \mu^{+}\mu^{-} e^{\pm}\nu_{e} \\ \nonumber
Z t\bar{t}\rightarrow \mu^{+}\mu^{-} e^{+}\nu_{e} b (e^{-}\bar{\nu}_{e} \bar{b}) \\ \nonumber
Zt\bar{b}\rightarrow \mu^{+}\mu^{-} e^{+}\nu_{e} \ b  \bar{b} \nonumber
\end{eqnarray}

For the event selection we impose the following criteria:
\begin{eqnarray}
&& \text{one electron (positron) with } p^{e}_{T} > 25\; \hbox{GeV}, \text{ and} \not\not\!\! {E_T} > 15\; \hbox{GeV}\ \label{eq:selection1a};\\  
&& \text{a pair of } \mu \text{ with } p^{\mu}_{T} > 25\; \hbox{GeV each},\ \label{eq:selection1b};\\  
&& \text{a pair of } \mu \text{ with } p^{\mu}_{T} > 25\; \hbox{GeV each} \text{ and reconstructed object } W^{\pm}. \label{eq:selection1c}
\end{eqnarray}

\begin{table*}[h!]    
\centering
\begin{tabular}{|c|c|c|c|c|}
\toprule
\hline
Process cross section (fb) & \makecell{Basic Selection \\ Eq.~\eqref{eq:basic_cuts}} & \makecell{Selection 1 \\ Eq.~\eqref{eq:selection1a}} & \makecell{Selection 2 \\ Eq.~\eqref{eq:selection1b}} & \makecell{Selection 3 \\ Eq.~\eqref{eq:selection1c}} \\
\hline
\hline
\midrule
$W^{\pm}\rightarrow \mu^{\pm} n^1_{3_L}(n^1_{4_L})$ & $2.28\times 10^{-2}$ & $1.54\times 10^{-2}$ & $2.7\times 10^{-3}$ & $1.7\times 10^{-3}$\\
\hline
$W^{\pm}Z, (W^{\pm}\rightarrow e^{\pm} \nu_{e} b,\ Z\rightarrow \mu^{+}\mu^{-})$ & $104.32$ & $71.82$ & $68.79$ & $47.51$\\
\hline
$t\bar{t}Z, (t^{(Wb)}\rightarrow e^{\pm} \nu_{e} b,\ Z\rightarrow \mu^{+}\mu^{-})$ & $0.3$ & $2.845\times 10^{-1}$ & $2.24\times 10^{-1}$ & $2.125\times 10^{-1}$ \\
\hline
$Ztb, (t^{(Wb)}\rightarrow e^{+} \nu_{e} b,\ Z\rightarrow \mu^{+}\mu^{-})$ & $3.98\times 10^{-2}$ & $2.92\times 10^{-2}$ & $2.88\times 10^{-2}$ & $2.13\times 10^{-2}$\\
\hline

\hline
\end{tabular}
\caption{Cross sections, in fb, for signal and background processes after successive selection criteria of Eqs.~\eqref{eq:basic_cuts} --\eqref{eq:selection1c}.}
\label{tab:sele_cuts}
\end{table*}

After we impose the selection criteria described in Eqs.~\eqref{eq:basic_cuts} --\eqref{eq:selection1c}, we are able to analyze the kinematics (dimension-full) and angular (dimension-less) observables from the final state particles produced by this channel. This analysis has the purpose of increase the significance of detecting $n^1_{3_L}(n^1_{4_L})$ at the next LHC run. We choose the following observables:

\begin{table*}[h!]    
\centering
\begin{tabular}{|c|c|c|c|}
\toprule
\hline
 & Dimension-full & \multicolumn{2}{|c|}{Dimensionless} \\
\hline
\hline
\midrule
\makecell{laboratory \\
referential \\
frame}  & \makecell{$M(\mu^{+},\mu^{-})$, $M(e,\mu^{+},\mu^{-})$ \\
                                 $M(n^1_{3_L}(n^1_{4_L}))$, $M_{T}(e^{\pm},\nu_{e})$, \\ 
                                 $p^{e_{1}}_{T}$, $p^{\mu_{1}}_{T}$, $p^{\mu_{2}}_{T}$, \\ $p^{n^1_{i_L}}_{T}$,$p^{W}_{T}$ } & 
                       \makecell{$\cos(\theta_{e,\not\not\!\! {\vec{E}_T}})$, $\cos(\theta_{\mu^{+},\mu^{-}})$,\\
                                 $\cos(\theta_{\mu_{1},W})$, $\cos(\theta_{\mu_{2},W})$, \\
                                 $\cos(\theta_{\mu_{1},n^1_{i_L}})$,
                                 $\cos(\theta_{\mu_{2},n^1_{i_L}})$, \\
                                 $\cos(\theta_{W,n^1_{i_L}})$, $\cos(\theta_{\mu_{1},\not\not\!\! {\vec{E}_T}})$, \\
                                 $\cos(\theta_{\mu_{1},\not\not\!\! {\vec{E}_T}})$} & 
                       \makecell{$\Delta R(\mu^{+},\mu^{-})$, $\Delta R(\mu_{1},W)$\\
                                 $\Delta R(\mu_{1},\not\not\!\! {\vec{E}_T})$, $\Delta R(\mu_{2},W)$, \\
                                 $\Delta R(\mu_{2},\not\not\!\! {\vec{E}_T})$, $\Delta R(e,\not\not\!\! {\vec{E}_T})$,\\
                                 $\Delta R(\mu_{1},n^1_{i_L})$, $\Delta R(\mu_{2},n^1_{i_L})$,\\
                                 $\Delta R(W,\not\not\!\! {\vec{E}_T})$}\\
\hline
\makecell{$n^{1}_{3_L} (n^{1}_{4_L})$ \\
referential \\
frame} & & \makecell{$\cos(\theta_{e,\not\not\!\! {\vec{E}_T}})_{n^{1}_{i_L}}$,
                                    $\cos(\theta_{\mu_{1},W})_{n^{1}_{i_L}}$,\\ $\cos(\theta_{\mu_{2},W})_{n^{1}_{i_L}}$, 
                                    $\cos(\theta_{\mu_{1},e})_{n^{1}_{i_L}}$,\\
                                    $\cos(\theta_{\mu_{2},e})_{n^{1}_{i_L}}$, 
                                    $\cos(\theta_{W,\not\not\!\! {\vec{E}_T}})_{n^{1}_{i_L}}$}  & 
                          \makecell{$\Delta R(\mu^{+},\mu^{-})_{n^{1}_{i_L}}$,
                                 $\Delta R(\mu_{1},W)_{n^{1}_{i_L}}$, \\
                                 $\Delta R(\mu_{1},\not\not\!\! {\vec{E}_T})_{n^{1}_{i_L}}$, 
                                 $\Delta R(\mu_{2},W)_{n^{1}_{i_L}}$, \\
                                 $\Delta R(\mu_{2},\not\not\!\! {\vec{E}_T})_{n^{1}_{i_L}}$, 
                                 $\Delta R(e,\not\not\!\! {\vec{E}_T})_{n^{1}_{i_L}}$}\\
\hline
\hline
\end{tabular}
\caption{Kinematic (Dimension-full) and angular (Dimension-less) observables selected to study the channel $p p \rightarrow \mu^{\pm}\mu^{\mp}e^{\pm}\nu_{e}$. We include dimensionless observables in two different referential frames: Center of Mass frame (top row) and $n^{1}_{i_L}$ rest frame (bottom row), where $\theta_{i,j}$ is the angle between the respective particles from either the final state or reconstructed objects, $W,n^{1}_{i_L}$, and $\Delta R (i,j)$ is the separation in the $\eta\times\phi$ plane defined by $\sqrt{(\Delta \phi)^{2} + (\Delta \eta)^{2}}$.}
\label{tab:vars}
\end{table*}

\begin{figure}[h!]
\begin{center}

\subfloat{%
\includegraphics[clip,scale=0.16]{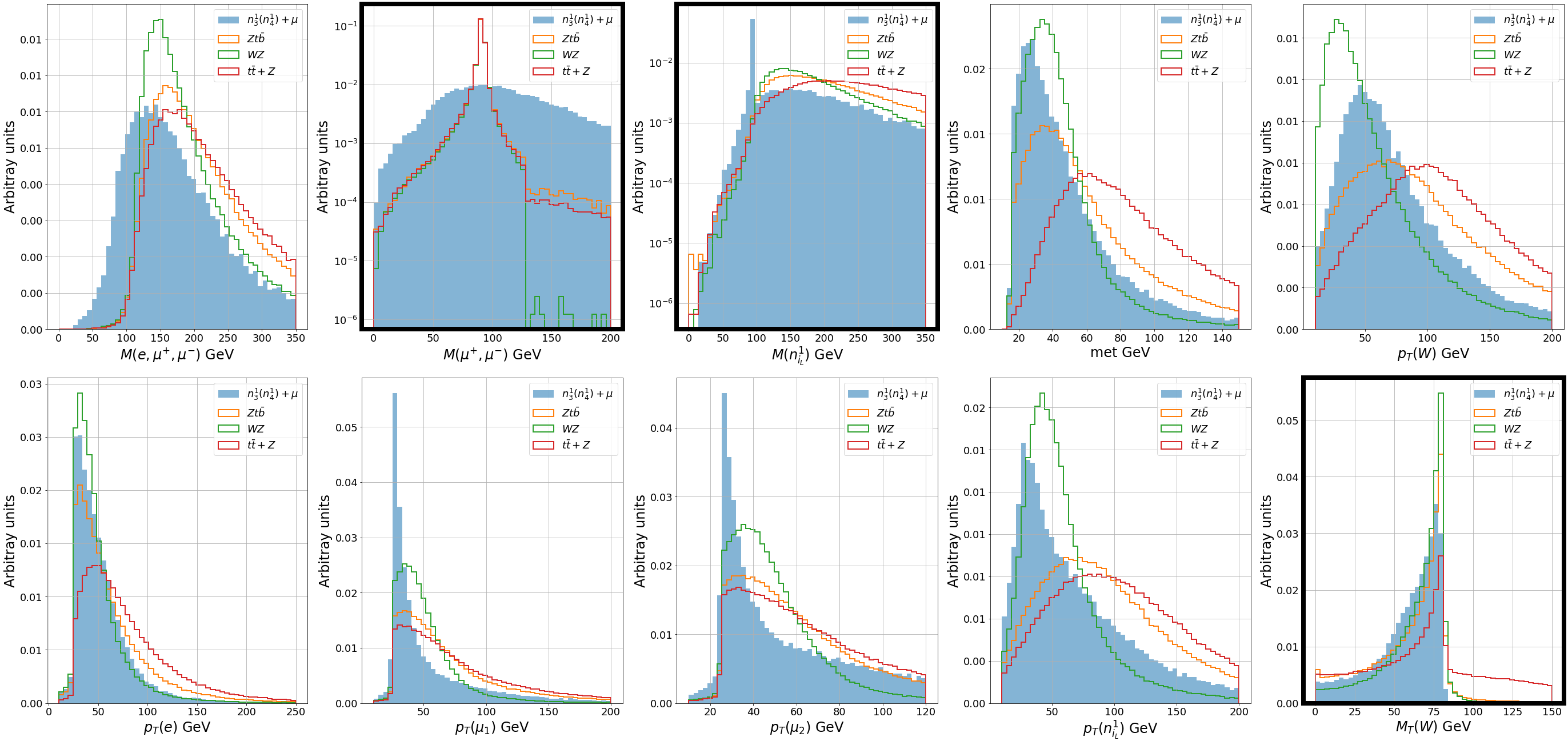}%
}

\caption{Kinematic (dimension-full) observables for the $p p \rightarrow \mu^{\pm}\mu^{\mp}e^{\pm}\nu_{e}$ channel. The blue region represents the kinematic distribution for the signal events, while the orange, green and red lines are the $Zt\bar{b}, WZ$ and $t\bar{t} + Z$ respective backgrounds. The $met$ variable corresponds to total missing transverse energy. We highlight the observables were the the signal plays a dominant role on the distributions.}
\label{fig_vars}
\end{center}
\end{figure}

\begin{figure}[h!]
\begin{center}

\subfloat{%
\includegraphics[clip,scale=0.22]{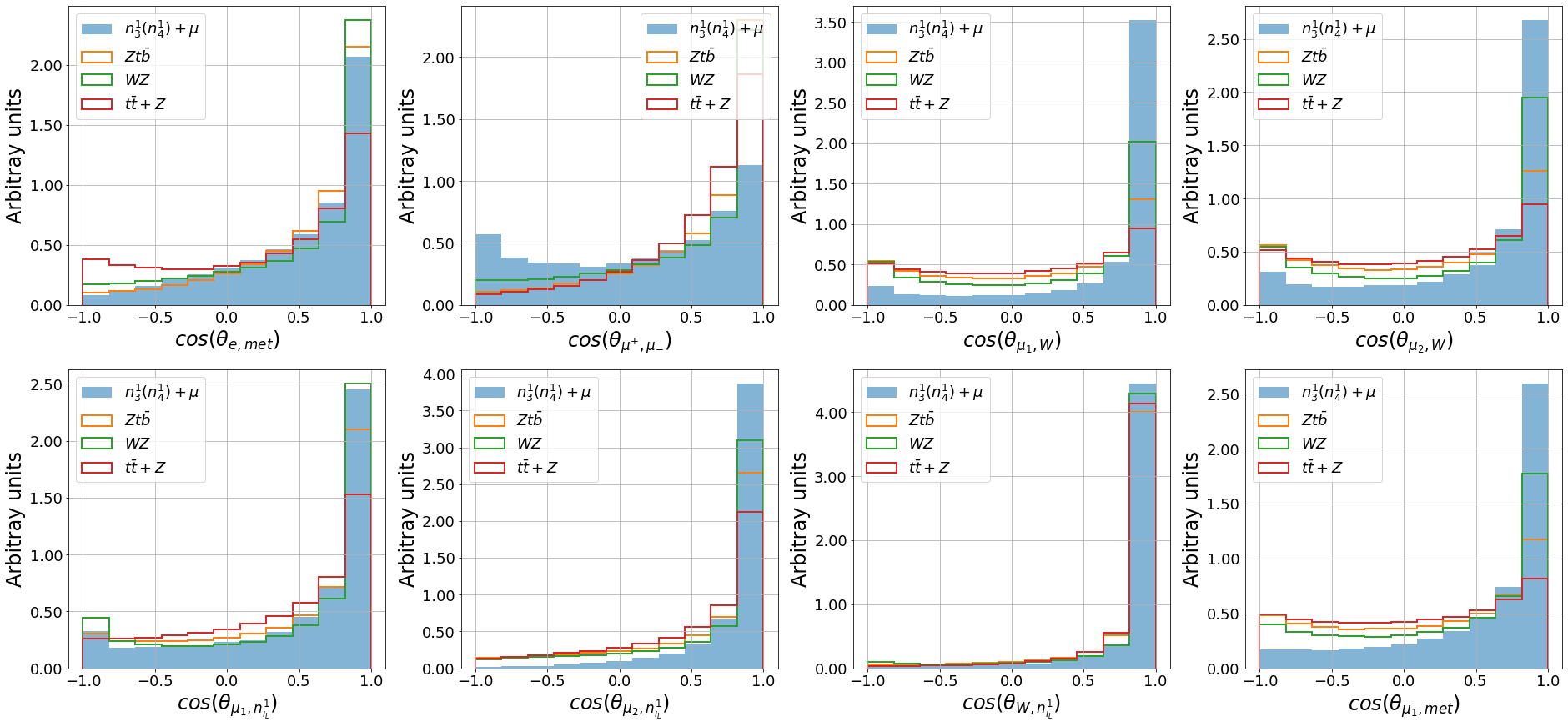}
}

\caption{Angular (dimensionless) observables for the $p p \rightarrow \mu^{\pm}\mu^{\mp}e^{\pm}\nu_{e}$ channel. Cosine of the angle between selected particles at the Center of mass and $n^1_{i,L}$ particle frames. The subscript indicate the observable is be taking in the $n^{1}_{3_L}(n^{1}_{4_L})$ reference frame. The blue region represents the kinematic distribution for the signal events, while the orange, green and red lines are the $Zt\bar{b}, WZ$ and $t\bar{t} + Z$ respective backgrounds. The $met$ variable corresponds to $\not\not\!\! {\vec{E}_T}$ vector direction.}
\label{fig_vars_cos}
\end{center}
\end{figure}

\begin{figure}[htp]
\begin{center}

\subfloat{%
\includegraphics[clip,scale=0.2]{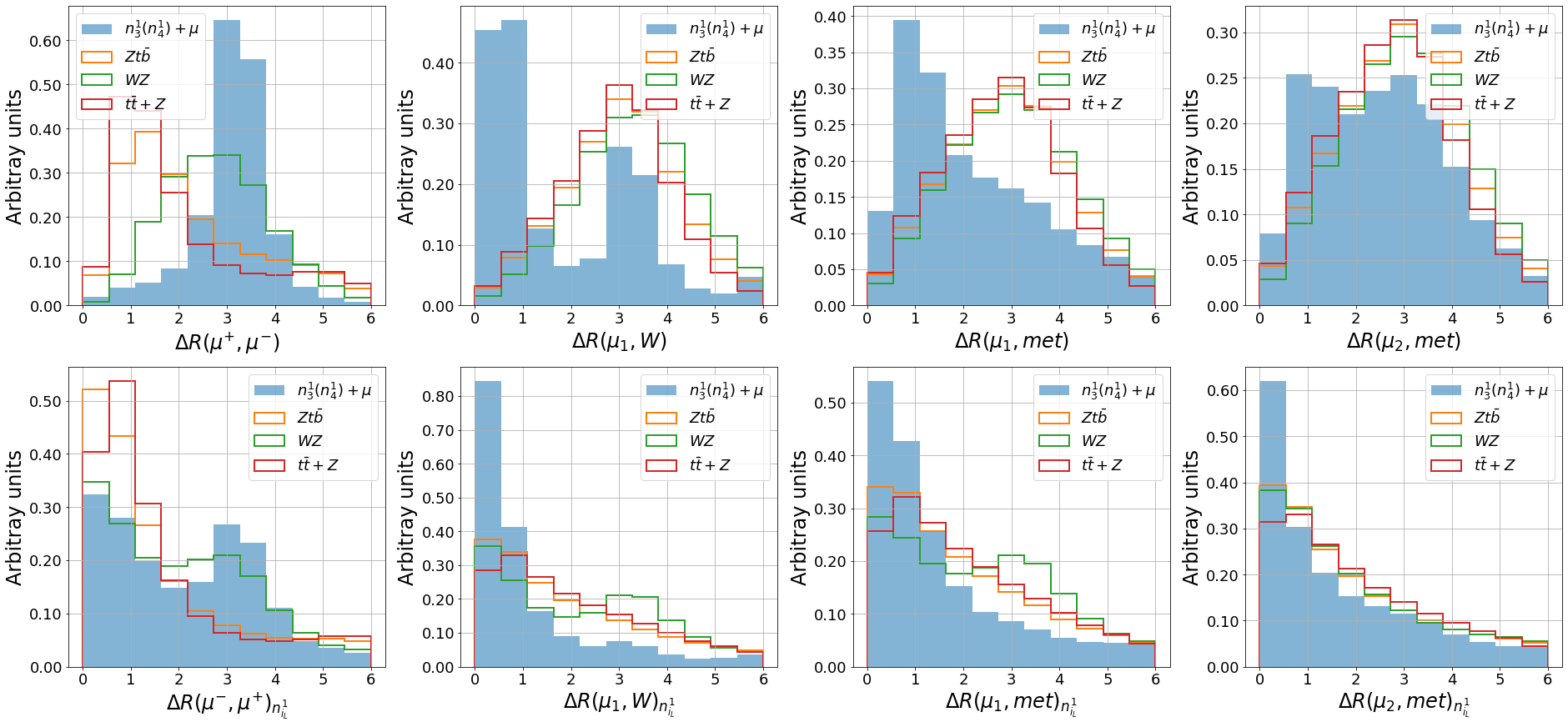}%
}

\caption{Angular (dimensionless) observables for the $p p \rightarrow \mu^{\pm}\mu^{\mp}e^{\pm}\nu_{e}$ channel. Separation in the $\eta \times \phi$ plane between selected particles  at the Center of mass and $n^1_{i,L}$ particle frames. The subscript script indicate the observable is be taking in the $n^{1}_{i_L}$ reference frame. The blue region represents the kinematic distribution for the signal events, while the orange, green and red lines are the $Zt\bar{b}, WZ$ and $t\bar{t} + Z$ respective backgrounds. The $met$ variable corresponds to $\not\not\!\! {\vec{E}_T}$ vector direction.}
\label{fig_vars_delta}
\end{center}
\end{figure}

In table ~\ref{tab:vars} we present the distributions for the observables of our analysis, and in  FIGs.~\ref{fig_vars} -- \ref{fig_vars_delta} we display the respective distributions. One naive approach is a simple cut and count analysis using the reconstructed  $n^1_{3_L}(n^1_{4_L})$ from the final state muon and reconstructed $W$ boson. However, due to the number of events for the background remained after the selection, even when we impose a cut window around the mass predicted for the $n^1_{3_L}(n^1_{4_L})$, buries completely our signal. To overcome this problem we make use of a Deep learning algorithm trained to distinguish the signal over the main irreducible background using the observables described before. We present the details of the architecture and training methodology in the section \ref{ML_tech}.
%%%%
%%%%
%%%%
%%%%

\clearpage
\subsection{$Z^{\prime}$ channel:}

\begin{figure}[h!]
\begin{center}
\subfloat{
\includegraphics[clip,scale=0.6]{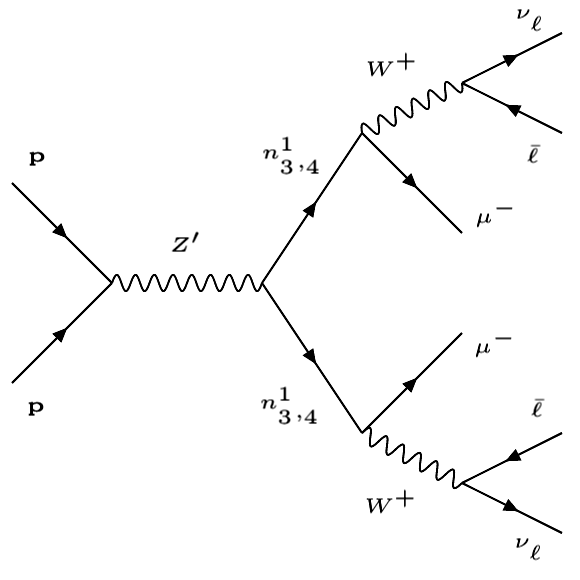}
}
\caption{Production of $n^1_{(3,4)}$ at the LHC via $Z^{\prime}$ channel.}
\label{Zchanell}
\end{center}
\end{figure}

Another production mechanism for the $n^{1}_{i_L}$ is through the production and subsequent  decay of $Z^{\prime}$, see FIG. (\ref{Zchanell}).
%%%%

%%%%
To investigate this channel we apply the same workflow where we generate  450000 events with 14 TeV and the same basic generation cuts described in Eq.~\eqref{eq:basic_cuts}. We then pass the hard scattering events through Pythia and Delphes to finally select the events based on the following selection criteria:

\begin{eqnarray}
&& \text{a pair of electron and positron with } p^{e}_{T} > 25\; \hbox{GeV}, \text{ and} \not\not\!\! {E_T} > 15\; \hbox{GeV}\;\label{eq:selection2a} \\  
&& \text{a pair of } \mu \text{ with } p^{\mu}_{T} > 25\; \hbox{GeV each},\;\label{eq:selection2b} \\  
&& \text{a pair of } \mu \text{ with } p^{\mu}_{T} > 25\; \hbox{GeV each } \text{and two reconstructed } W^{\pm}. \label{eq:selection2c} 
\end{eqnarray}

\begin{table*}[h!]    
\centering
\begin{tabular}{|c|c|c|c|c|}
\toprule
\hline
Process cross section (fb) & \makecell{Basic Selection \\ Eq.~\eqref{eq:basic_cuts}} & \makecell{Selection 1 \\ Eq.~\eqref{eq:selection2a}} & \makecell{Selection 2 \\ Eq.~\eqref{eq:selection2b}} & \makecell{Selection 3 \\ Eq.~\eqref{eq:selection2c}} \\
\hline
\hline
\midrule
$Z'\rightarrow n^1_{3_L}\bar{n}^1_{3_L}$ & $4.32\times 10^{-2}$ & $3.88\times 10^{-2}$ & $3.8\times 10^{-2}$ & $3.36\times 10^{-2}$\\
\hline
$W^{+}W^{-}Z, (W^{\pm}\rightarrow e^{\pm} \nu_{e} b,\ Z\rightarrow \mu^{+}\mu^{-})$ & $4.0\times 10^{-2}$ & $2.7\times 10^{-2}$ & $2.52\times 10^{-2}$ & $1.42\times 10^{-2}$\\
\hline
$t\bar{t}Z, (t^{(Wb)}\rightarrow e^{\pm} \nu_{e} b,\ Z\rightarrow \mu^{+}\mu^{-})$ & $3.0\times 10^{-1}$ & $2.234\times 10^{-1}$ & $1.81\times 10^{-1}$ & $1.24\times 10^{-1}$ \\
\hline

\hline
\end{tabular}
\caption{Cross sections, in fb, for signal and background processes after successive selection criteria of Eqs.~\eqref{eq:basic_cuts} --\eqref{eq:selection2c}.}
\label{tab:sele_cuts_zp}
\end{table*}

\begin{figure}[h!]

\centering
    \subfloat[]{%
    \includegraphics[width=.5\linewidth]{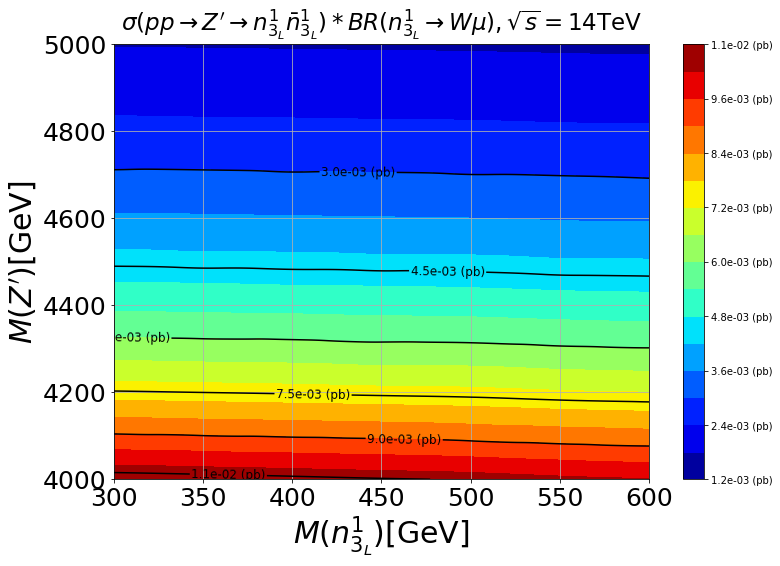}%
    \includegraphics[width=.5\linewidth]{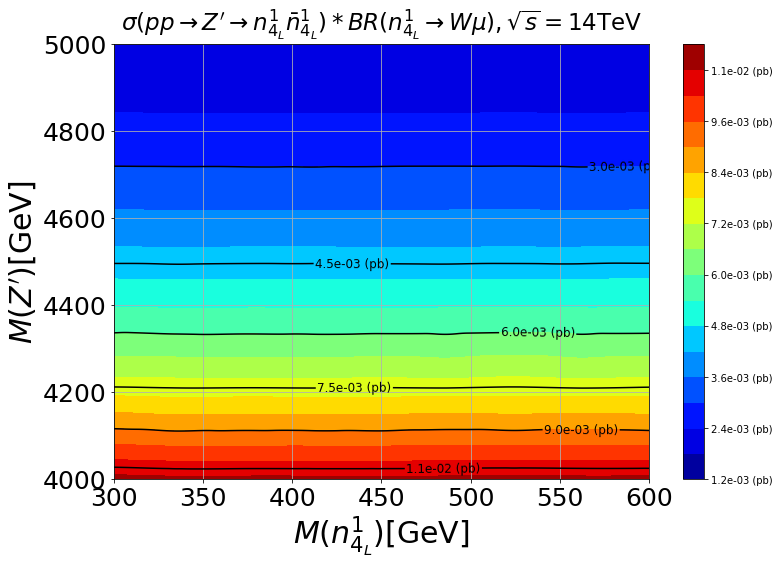}%
    }
 
\caption{Cross section times branching ratio dependency of the processes $pp \rightarrow Z' \rightarrow n^{1}_{3L} \bar{n}^{1}_{3L}$ (left) and $pp \rightarrow Z' \rightarrow n^{1}_{4L} \bar{n}^{1}_{4L}$ (right) for the masses of the Z' (y-axis)  and $n^{1}_{3L} (n^{1}_{4L})$.}
\label{zp_xsec}
\end{figure}

The $W^{\pm}$ bosons are reconstructed from the final state electrons and the $\not\not\!\! {E_T}$. In our simulations we set the value for the $Z^{\prime}$ mass to 4 TeV and $n^{1}_{3L} (n^{1}_{4L})$ to 96.31 GeV which are consistent with the current estimate limits \cite{Sirunyan:2018exx,Aad:2019fac} for the expected $Z^{\prime}$ mass. In FIGs.~\ref{zp_xsec} we display the cross section for a given range of $Z
^{\prime}$ mass against the $n^{1}_{i_L}$ ones. The region explored in this paper offers a sizeable cross section for the production of a $Z^{\prime}$ and its subsequent decay into $n^{1}_{i_L}$. 

For the main irreducible background we have:

\begin{itemize}
    \item $Zt\bar{t}\rightarrow \mu^{+}\mu^{-} e^{+}\nu_{e} \ b e^{-}\bar{\nu}_{e} \bar{b}$
    \item $ZW^{+}W^{-}\rightarrow \mu^{+}\mu^{-} e^{+}\nu_{e} e^{-}\bar{\nu}_{e}$
\end{itemize}

%%%%

\begin{figure}[h!]
\begin{center}

\subfloat{%
\includegraphics[clip,scale=0.22]{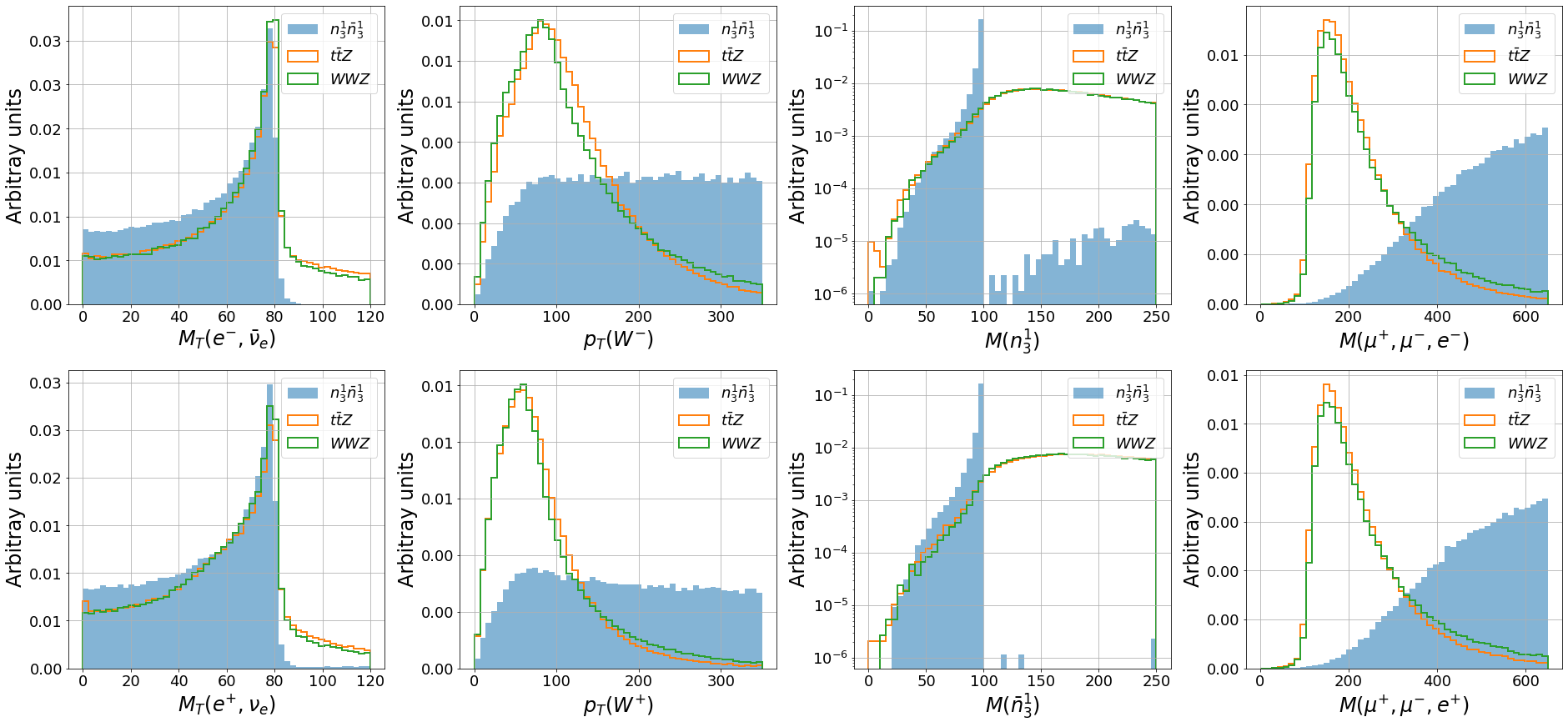}%
}

\caption{Kinematic (dimension-full) observables for the $p p \rightarrow \mu^{\pm}\mu^{\mp}e^{\pm}\nu_{e} e^{\mp}\nu_{e}$ channel. The blue region represents the kinematic distribution for the signal events, while the orange and green lines are the $t\bar{t}Z$ and $WWZ$ backgrounds.}
\label{fig_vars3}
\end{center}
\end{figure}

\begin{figure}[h!]
\begin{center}

\subfloat[cosine of the angle between selected particles.]{%
\includegraphics[clip,scale=0.29]{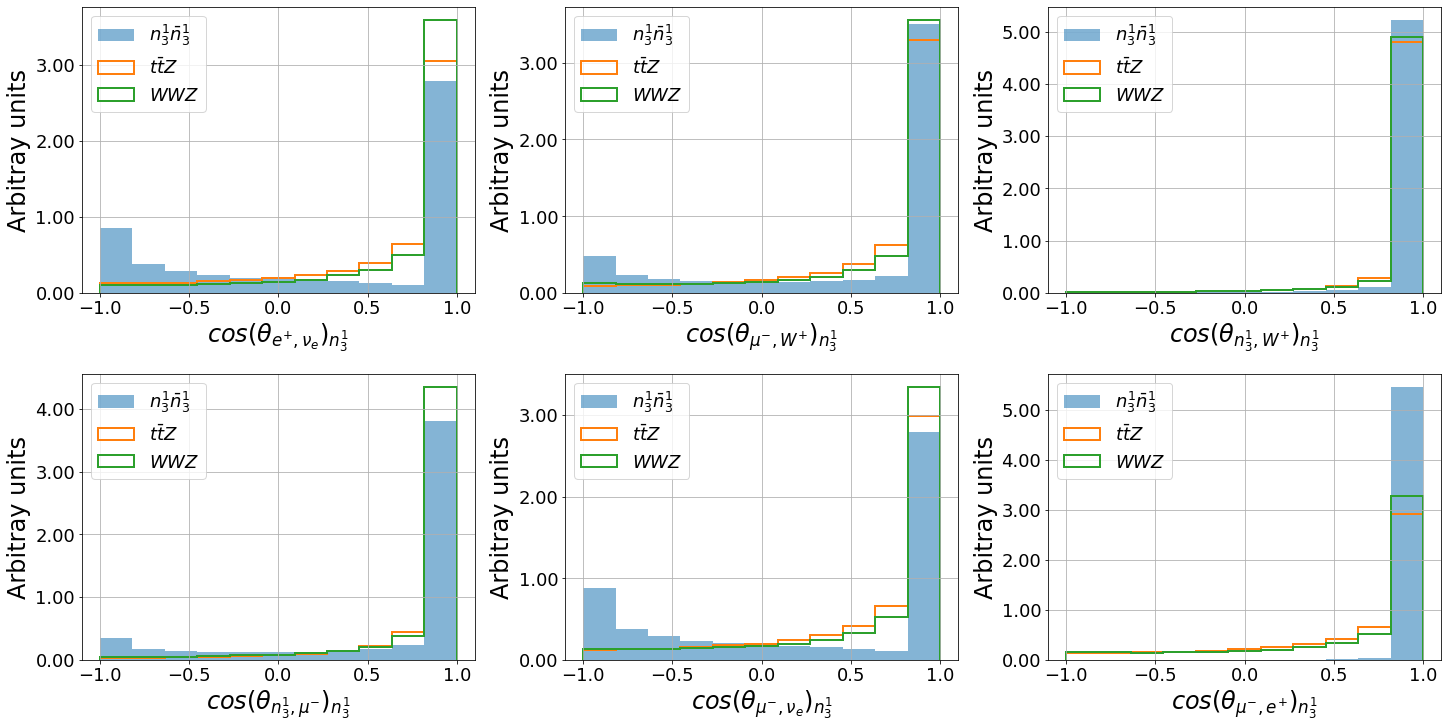}%
}

\subfloat[separation between selected particles.]{%
\includegraphics[clip,scale=0.29]{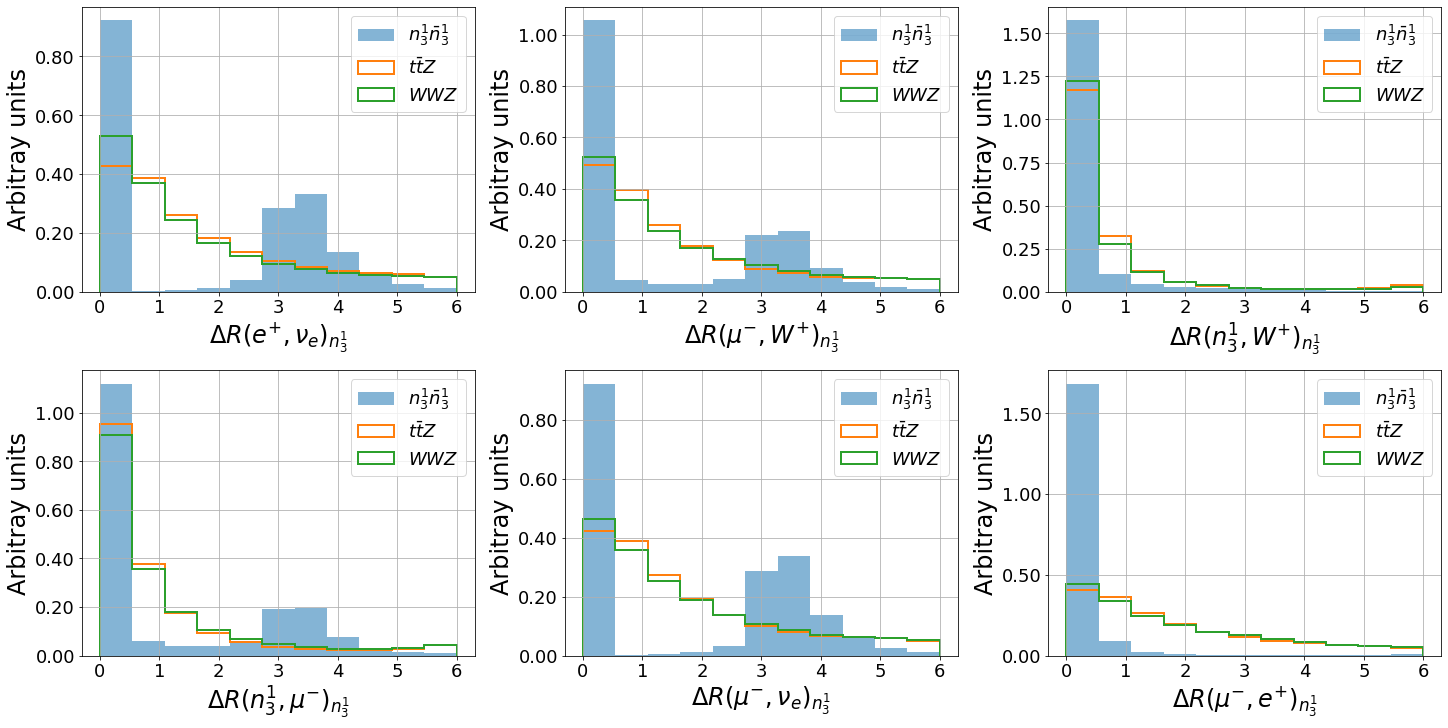}%
}

\caption{Angular (dimensionless) observables for the $p p \rightarrow \mu^{\pm}\mu^{\mp}e^{\pm}\nu_{e} e^{\mp}\nu_{e}$ channel. The blue region represents the angular distribution of our signal, while the orange and green lines are the $t\bar{t}Z$ and $WWZ$ backgrounds. The subscript script indicate the observable is be taking in the $n^{1}_{i_L}$ object reconstructed reference frame.}
\label{fig_vars4}
\end{center}
\end{figure}

\begin{figure}[h!]
\begin{center}

\subfloat[cosine of the angle between selected particles.]{%
\includegraphics[clip,scale=0.29]{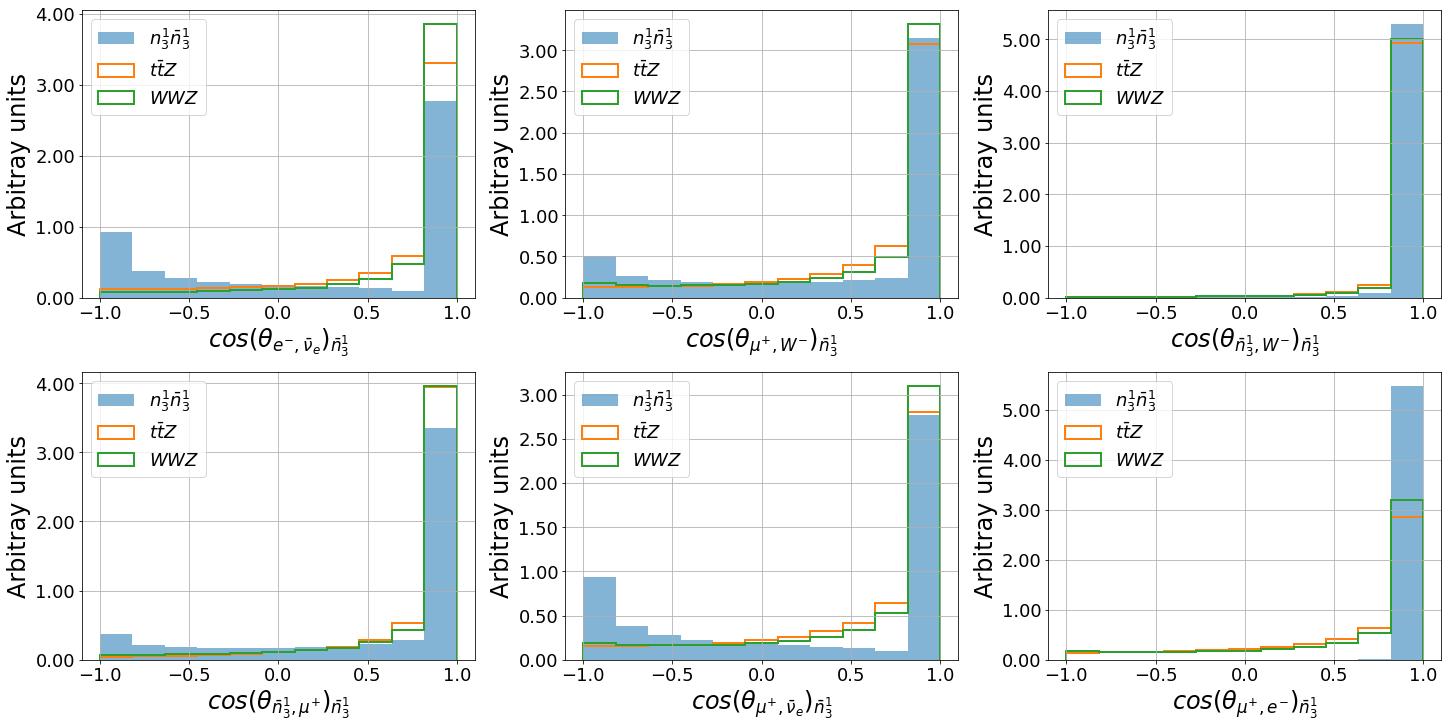}%
}

\subfloat[separation between selected particles.]{%
\includegraphics[clip,scale=0.29]{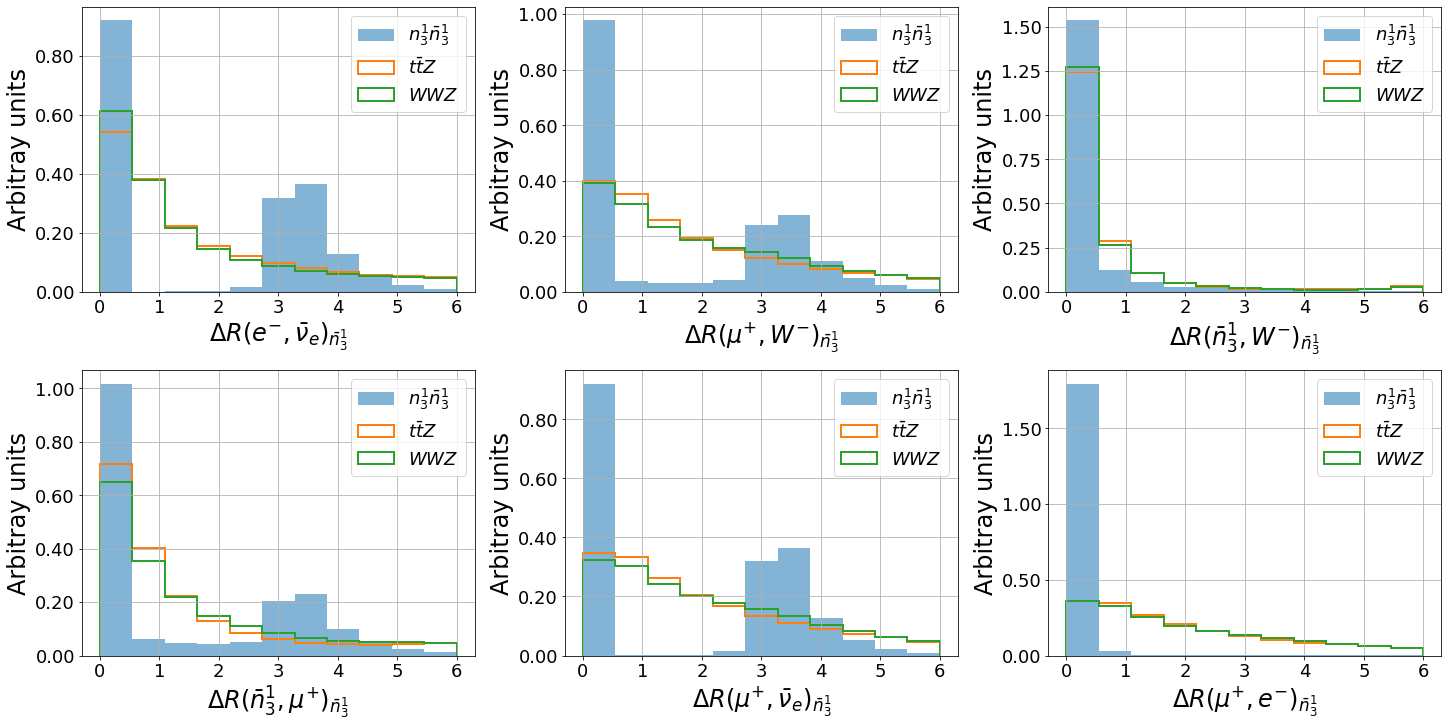}%
}

\caption{Angular (dimensionless) observables for the $p p \rightarrow \mu^{\pm}\mu^{\mp}e^{\pm}\nu_{e} e^{\mp}\nu_{e}$ channel. The subscript script indicate the observable is be taking in the $n^{1}_{i_L}$ object reconstructed reference frame.}
\label{fig_vars5}
\end{center}
\end{figure}

This channel contains six leptons as final state particles, 4 visible ($\mu^{+}, \mu^{-}, e^{+}, e^{-}$) and 2 invisible ($\nu_{e},\bar{\nu}_{e}$), which opens up the number of observables we can use to distinguish the signal over background. We choose the following dimension-full and dimensionless variables, see TABLE IV, and in  FIGs. \ref{fig_vars3} -- \ref{fig_vars5} we display the respective distributions.

\begin{table*}[h!]    
\centering
\begin{tabular}{|c|c|c|c|}
\toprule
\hline
 & Dimension-full & \multicolumn{2}{|c|}{Dimensionless} \\
\hline
\hline
\midrule
\makecell{laboratory \\
referential \\
frame}  & \makecell{$M_{T}(e^{-},\not\not\!\! {E_T})$, $M_{T}(e^{+},\not\not\!\! {E_T})$, \\
                    $p_{T}(\mu^{-})$,$p_{T}(\mu^{+})$, \\
                    $p_{T}(e^{-})$, $p_{T}(e^{+})$, \\
                    $p_{T}(W^{+})$,$p_{T}(W^{-})$, \\
                    $p_{T}(n^{1}_{iL})$, $p_{T}(\bar{n}^{1}_{iL})$, \\
                    $\not\not\!\! {E_T}$, $M(n^{1}_{iL})$, \\
                    $M(\bar{n}^{1}_{iL})$, $M(\mu^{+},\mu^{-},e^{+})$, \\
                    $M(\mu^{+},\mu^{-})$, $M(e^{+},e^{-})$} & 
         \makecell{$\cos(\theta_{e^{-},\not\not\!\! {\vec{E}_T}})$, $\cos(\theta_{e^{+},\not\not\!\! {\vec{E}_T}})$,\\
                 $\cos(\theta_{e^{-},e^{+}})$, $\cos(\theta_{\mu^{-},W^{+}})$, \\
                 $\cos(\theta_{\mu^{-},e^{+}})$, $\cos(\theta_{\mu^{-},\not\not\!\! {\vec{E}_T}})$, \\
                 $\cos(\theta_{\mu^{+},W^{-}})$, $\cos(\theta_{\mu^{+},e^{+}})$, \\
                 $\cos(\theta_{\mu^{+},\not\not\!\! {\vec{E}_T}})$, $\cos(\theta_{n^{1}_{iL},W^{+}})$, \\
                 $\cos(\theta_{n^{1}_{iL},\mu^{-}})$, $\cos(\theta_{n^{1}_{iL},e^{+}})$, \\
                 $\cos(\theta_{\bar{n}^{1}_{iL},W^{-}})$, $\cos(\theta_{\bar{n}^{1}_{iL},\mu^{+}})$,\\
                 $\cos(\theta_{\bar{n}^{1}_{iL},e^{-}})$, $\cos(\theta_{W^{+},W^{-}})$, \\
                 $\cos(\theta_{\mu^{+},\mu^{-}})$, $\cos(\theta_{e^{+},e^{-}})$} & 
        \makecell{$\Delta R(e^{-},\not\not\!\! {\vec{E}_T})$, $\Delta R(e^{+},\not\not\!\! {\vec{E}_T})$, \\
        $\Delta R(e^{-},e^{+})$, $\Delta R(\mu^{-},W^{+})$, \\
        $\Delta R(\mu^{-},e^{+})$, $\Delta R(\mu^{-},\not\not\!\! {\vec{E}_T})$, \\
        $\Delta R(\mu^{+},W^{-})$, $\Delta R(\mu^{+},e^{+})$, \\
        $\Delta R(\mu^{+},\not\not\!\! {\vec{E}_T})$, $\Delta R(n^{1}_{iL},W^{+})$, \\
        $\Delta R(n^{1}_{iL},\mu^{-})$, $\Delta R(n^{1}_{iL},e^{+})$, \\
        $\Delta R(\bar{n}^{1}_{iL},W^{-})$, $\Delta R(\bar{n}^{1}_{iL},\mu^{+})$, \\
        $\Delta R(\bar{n}^{1}_{iL},e^{-})$, $\Delta R(W^{+},W^{-})$, \\
        $\Delta R(\mu^{+},\mu^{-})$, $\Delta R(e^{+},e^{-})$}\\
\hline
\makecell{$n^{1}_{i_L}$ \\
referential \\
frame} & & \makecell{
$\cos(\theta_{e^{+},\not\not\!\! {\vec{E}_T}})_{n^{1}_{iL}}$, $cos(\theta_{\mu^{-},W^{+}})_{n^{1}_{iL}}$, \\
$\cos(\theta_{n^{1}_{iL},W^{+}})_{n^{1}_{iL}}$, $\cos(\theta_{n^{1}_{iL},\mu^{-}})_{n^{1}_{iL}}$, \\
$\cos(\theta_{\mu^{-},\not\not\!\! {\vec{E}_T}})_{n^{1}_{iL}}$, $cos(\theta_{\mu^{-},e^{+}})_{n^{1}_{iL}}$
 }  & 
           \makecell{ 
$\Delta R(e^{+},\not\not\!\! {\vec{E}_T})_{n^{1}_{iL}}$, $\Delta R(\mu^{-},W^{+})_{n^{1}_{iL}}$, \\
$\Delta R(n^{1}_{iL},W^{+})_{n^{1}_{iL}}$, $\Delta R(n^{1}_{iL},\mu^{-})_{n^{1}_{iL}}$, \\
$\Delta R(\mu^{-},\not\not\!\! {\vec{E}_T})_{n^{1}_{iL}}$, $\Delta R(\mu^{-},e^{+})_{n^{1}_{iL}}$
           }\\
\hline
\makecell{$\bar{n}^{1}_{i_L}$ \\
referential \\
frame} & & \makecell{
$\cos(\theta_{e^{-},\not\not\!\! {\vec{E}_T}})_{\bar{n}^{1}_{iL}}$, $\cos(\theta_{\mu^{+},W^{-}})_{\bar{n}^{1}_{iL}}$, \\
$\cos(\theta_{\bar{n}^{1}_{iL},W^{-}})_{\bar{n}^{1}_{iL}}$, $\cos(\theta_{\bar{n}^{1}_{iL},\mu^{+}})_{\bar{n}^{1}_{iL}}$, \\
$\cos(\theta_{\mu^{+},\not\not\!\! {\vec{E}_T}})_{\bar{n}^{1}_{iL}}$, $\cos(\theta_{\mu^{+},e^{-}})_{\bar{n}^{1}_{iL}}$
 }  & 
           \makecell{ 
           $\Delta R(e^{-},\not\not\!\! {\vec{E}_T})_{\bar{n}^{1}_{iL}}$, $\Delta R(\mu^{+},W^{-})_{\bar{n}^{1}_{iL}}$, \\
           $\Delta R(\bar{n}^{1}_{iL},W^{-})_{\bar{n}^{1}_{iL}}$, 
           $\Delta R(\bar{n}^{1}_{iL},\mu^{+})_{\bar{n}^{1}_{iL}}$, \\
           $\Delta R(\mu^{+},\not\not\!\! {\vec{E}_T})_{\bar{n}^{1}_{iL}}$, 
           $\Delta R(\mu^{+},e^{-})_{\bar{n}^{1}_{iL}}$
           }\\           
\hline
\hline
\end{tabular}
\caption{Kinematic (Dimension-full) and angular (Dimension-less) observables selected to study the channel . We include dimensionless observables in three different referential frames: Center of Mass frame (top row), $n^{1}_{i_L}$ rest frame (middle row) and $\bar{n}^{1}_{i_L}$ rest frame (bottom row), where $\theta_{i,j}$ is the angle between the respective particles from either the final state or reconstructed objects, $W,n^{1}_{i_L}$, and $\Delta R (i,j)$ is the separation in the $\eta\times\phi$ plane defined by $\sqrt{(\Delta \phi)^{2} + (\Delta \eta)^{2}}$.}
\label{tab:vars_Zp}
\end{table*}

%%%%

%%%%

%%%%

\clearpage
\subsection{Deep learning analysis: Methods and results}
\label{ML_tech}
After we select the events and gather the kinematic and angular information we can feed this information into a Neural Network (NN) designed to proper separate signal over background. Due to the simplicity of the data-set of our events, which store the information from the events as tables where each row correspond to an event entry and the columns are the observables, we decide to work with a fully connected NN. However, we still have to choose some important parameters for the NN: number of layers, number of neurons, kernel initializes, etc. The decision of choose the correct parameters directly reflect the efficiency of our NN, which can be translated into significance of discovery, or not, of the particles predicted by the model. 

This selection  is often refereed as hyperparameter optimization. A first approach is to use "brute force" to tune the hyperparameters by using a grid search,  but the number of combinations and the computational time to test each one of them increases exponentially. More efficient ways beyond grid search are random sampling or using gaussian process algorithms to learn the best hyperparameters. Another way to tackle this problem is to use genetic/evolutionary algorithms, as in Ref.~\cite{Freitas:2019hbk}. 

To test the different architectures, as well the modifications and fining tuning of the parameters, we set up an evolutionary algorithm to test the different combinations of parameters by creating a set of populations. In our case we restrict the population to 25 models, and keep the top 5 models with highest accuracy, after 5 rounds (generations) we obtain the top 3 architectures sorted by accuracy and we select the best one to continue our analysis. This full process takes around 2 hours in a NVIDIA GTX 1070 GPU. We use Tensorflow 2.0 \cite{Abadi:2016kic} to build, train and evaluate our models.

The best architecture and hyperparameters found by our genetic algorithm consist of a 5 layers NN each one with 512 neurons with a Rectified Linear Unit (a.k.a. ReLU) activation function with the exception of the top layers which consist of a layer with 4 neurons, one for each channel analysed ($\mu \ n^1_{3_L}(\mu \ n^1_{4_L})$, $Zt\bar{b}, WZ, t\bar{t}Z$), and a \textit{sigmoid} as activation function. We also found that initial random weights for the layers sampled from normal distribution and L2 regularization with a value of $10^{-7}$ gives the best significance. We also found a similar architecture for the channel $n^1_{3_L}\bar{n}^1_{3_L}(n^1_{4_L}\bar{n}^1_{4_L})$, with the only difference that at the top layer we have 3 neurons, one for each channel ($n^1_{3_L}\bar{n}^1_{3_L}(n^1_{4_L}\bar{n}^1_{4_L}), t\bar{t}Z, WWZ$). 

Our data sets consist of tables where each row corresponds to an event entry and the columns are the kinematics and angular distributions we described in the sections. Due to the selection criteria \ref{tab:sele_cuts} and \ref{tab:sele_cuts_zp} we impose into the the signal and backgrounds events, we ended up with an imbalanced number of events for each channel, this can lead the DNN model to over-fit towards the majority class, which turns the model unable to make correct predictions for the classes we are interested. To overcome this problem we balance the original data set using Synthetic Minority Over-sampling Technique (SMOTE) \cite{abs-1106-1813}, we first dived the original data set into 80\% to generate the balance data set and 20\% to use our validation set.

\begin{table*}[ht!]
\centering
\begin{tabular}{c|c|c|c}
\toprule
\hline
    Process & Original & Training (SMOTE) & Test/Validation  \\
    \hline\hline
     $W^{\pm}\rightarrow \mu^{\pm} n^1_{3_L}(n^1_{4_L})$ & (32060, 41) & (254528, 41) & (6243, 41)  \\
    \hline
    $W^{\pm}Z, (W^{\pm}\rightarrow e^{\pm} \nu_{e} b,\ Z\rightarrow \mu^{+}\mu^{-})$ & (205162, 41) & (254528, 41) & (41019, 41)  \\
    \hline
    $t\bar{t}Z, (t^{(Wb)}\rightarrow e^{\pm} \nu_{e} b,\ Z\rightarrow \mu^{+}\mu^{-})$ & (318443, 41) & (254528, 41) & (63915, 41)  \\
    \hline
    $Ztb, (t^{(Wb)}\rightarrow e^{+} \nu_{e} b,\ Z\rightarrow \mu^{+}\mu^{-})$ & (240033, 41) & (254528, 41) & (47963, 41)  \\
    \hline
    total (channel 1) & (890011, 41) & (1018112, 41) &  (159140, 41)\\
    \hline\hline
    $Z'\rightarrow n^1_{3_L}\bar{n}^1_{3_L}\ (n^1_{4_L}\bar{n}^1_{4_L})$ & (350140, 77) & (279963, 77) & (70177, 77) \\
    \hline
    $W^{+}W^{-}Z, (W^{\pm}\rightarrow e^{\pm} \nu_{e} b,\ Z\rightarrow \mu^{+}\mu^{-})$ & (159303, 77) & (279963, 77) & (32023, 77) \\
    \hline
    $t\bar{t}Z, (t^{(Wb)}\rightarrow e^{\pm} \nu_{e} b,\ Z\rightarrow \mu^{+}\mu^{-})$ & (185562, 77) & (279963, 77) & (36801, 77) \\
    \hline
    total (channel 2) & (695005, 77) & (839889, 77) &  (139001, 77)\\
    \hline\hline
\end{tabular}
\caption{Our data set. The first dimension corresponds to the number of events entries of each channel and the second dimension is the number of features (i.e. kinematic and angular variables). The first column shows the number of events survived after we apply the selection cuts Eq. \eqref{eq:selection2a} -- Eq. \eqref{eq:selection2c}, the \textbf{Training (SMOTE)} shows the balanced data sets after we apply the SMOTE algorithm to 80\% of the original events. The \textbf{Test/Validation} sets are the remain 20\% of the original selected events.} 
\label{tab:dataset}
\end{table*}

%%%%
\begin{figure}[h!]
\begin{center}

\subfloat[Signal efficiency over background rejections (left) and NN scores for the signal and backgrounds classes (right) for $\mu \ n^1_{3_L}(\mu \ n^1_{4_L})$.]{%
\includegraphics[clip,scale=0.37]{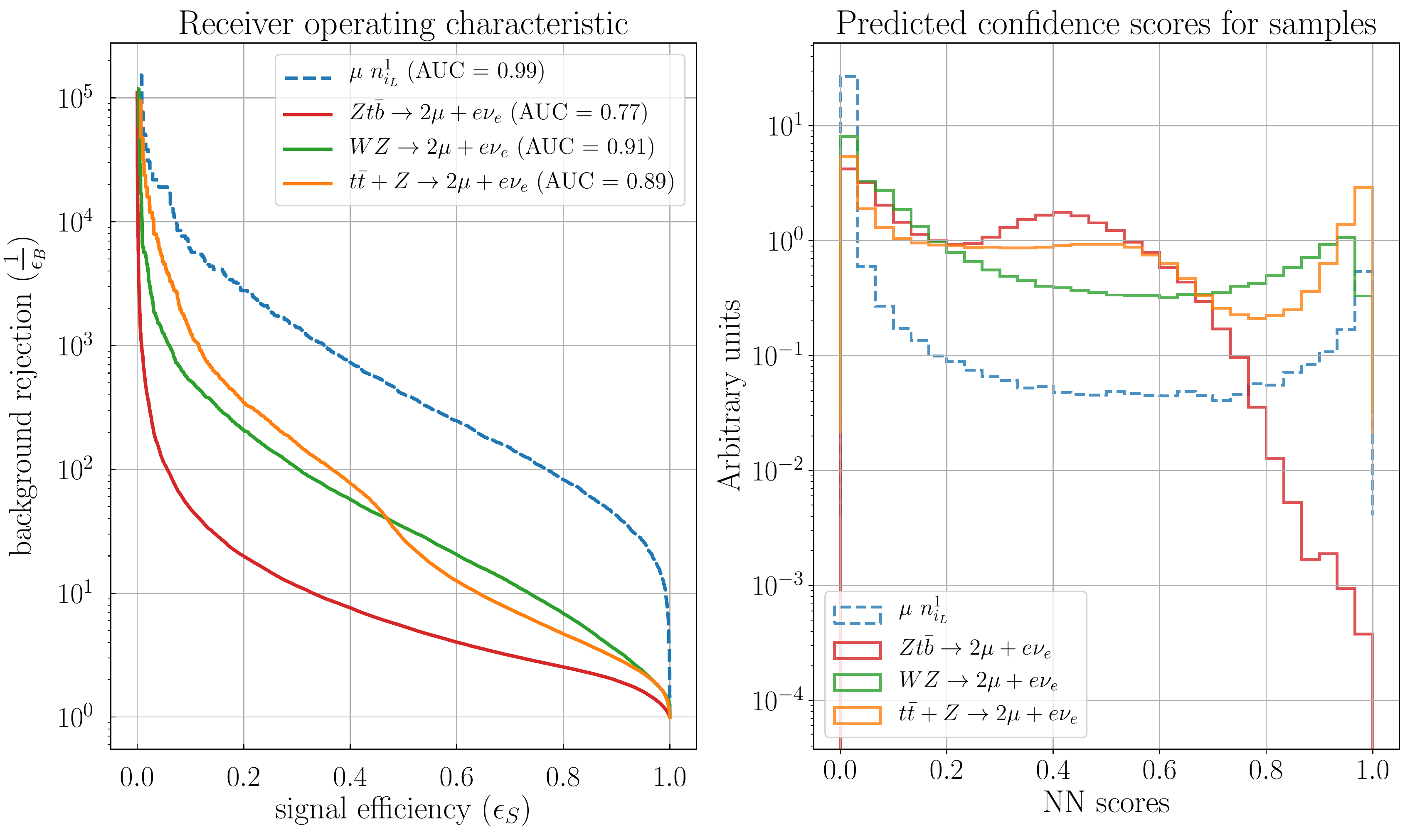}%
}

\subfloat[Signal efficiency over background rejections (left) and NN scores for the signal and backgrounds classes (right) for $n^1_{3_L}\bar{n}^1_{3_L}(n^1_{4_L}\bar{n}^1_{4_L})$]{%
\includegraphics[clip,scale=0.37]{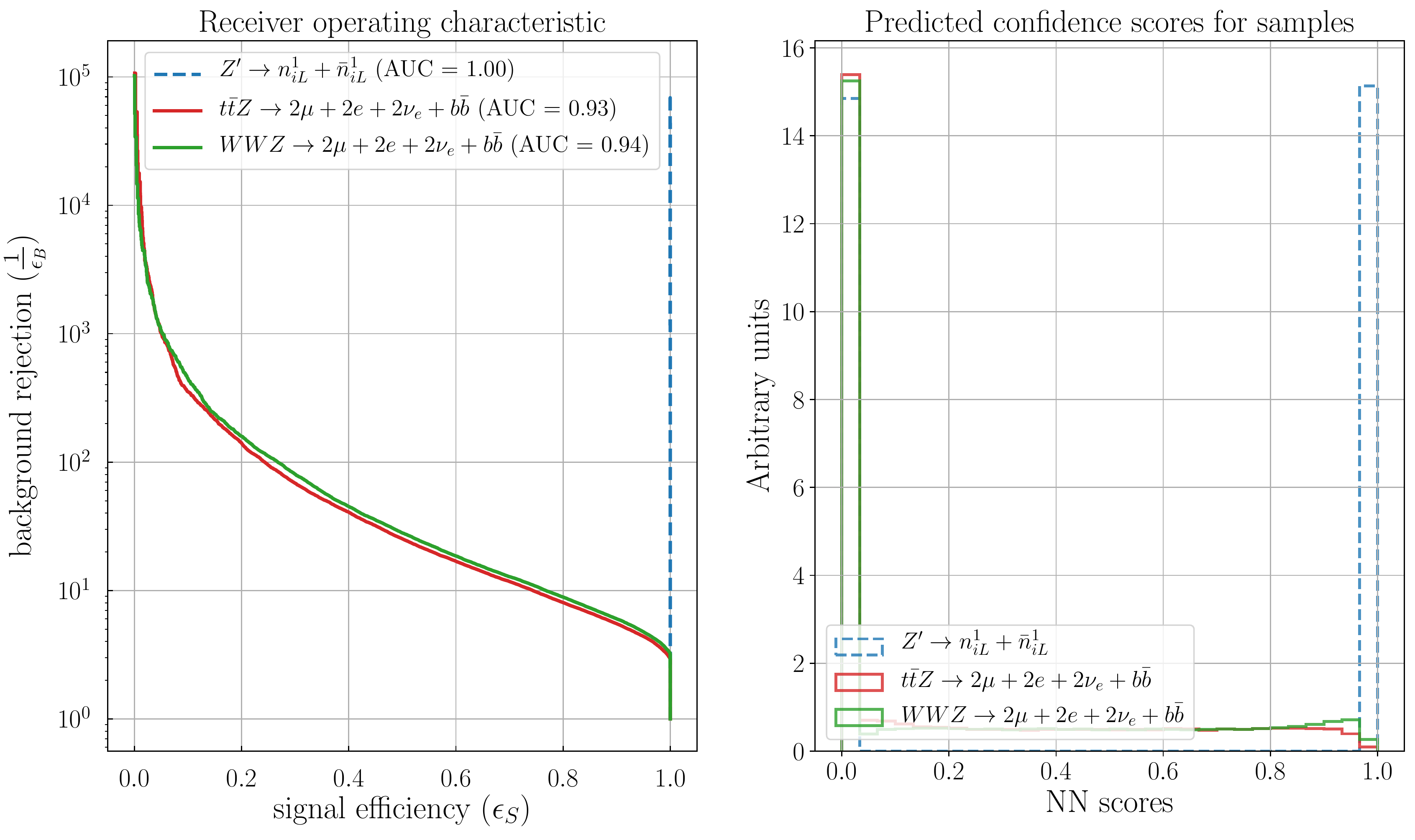}%
}

\caption{Signal efficiency over background rejections and prediction scores assigned by the Neural Network for the signal channel $p p \rightarrow \mu^{\pm}\mu^{\mp}e^{\pm}\nu_{e}$ (a) and $p p \rightarrow \mu^{\pm}\mu^{\mp}e^{\pm}\nu_{e} e^{\mp}\nu_{e}$ (b) and their respective backgrounds.}
\label{ml_res}
\end{center}
\end{figure}

We can evaluate the performance of our NN by look into the signal efficiency over the background rejection. The left panel of Fig.~\eqref{ml_res} show the signal efficiency and the background rejection for both channels analysed while the right panel gives us the normalized number of entries for a given NN prediction score. A simple figure to evaluate how good is the signal-background separation is the area under the ROC curve, AUC. The closer AUC is to one, the better we should expect the backgrounds can be cleaned up for a giving signal efficiency.

\begin{figure}[h!]
\begin{center}

\subfloat{%
\includegraphics[clip,scale=0.37]{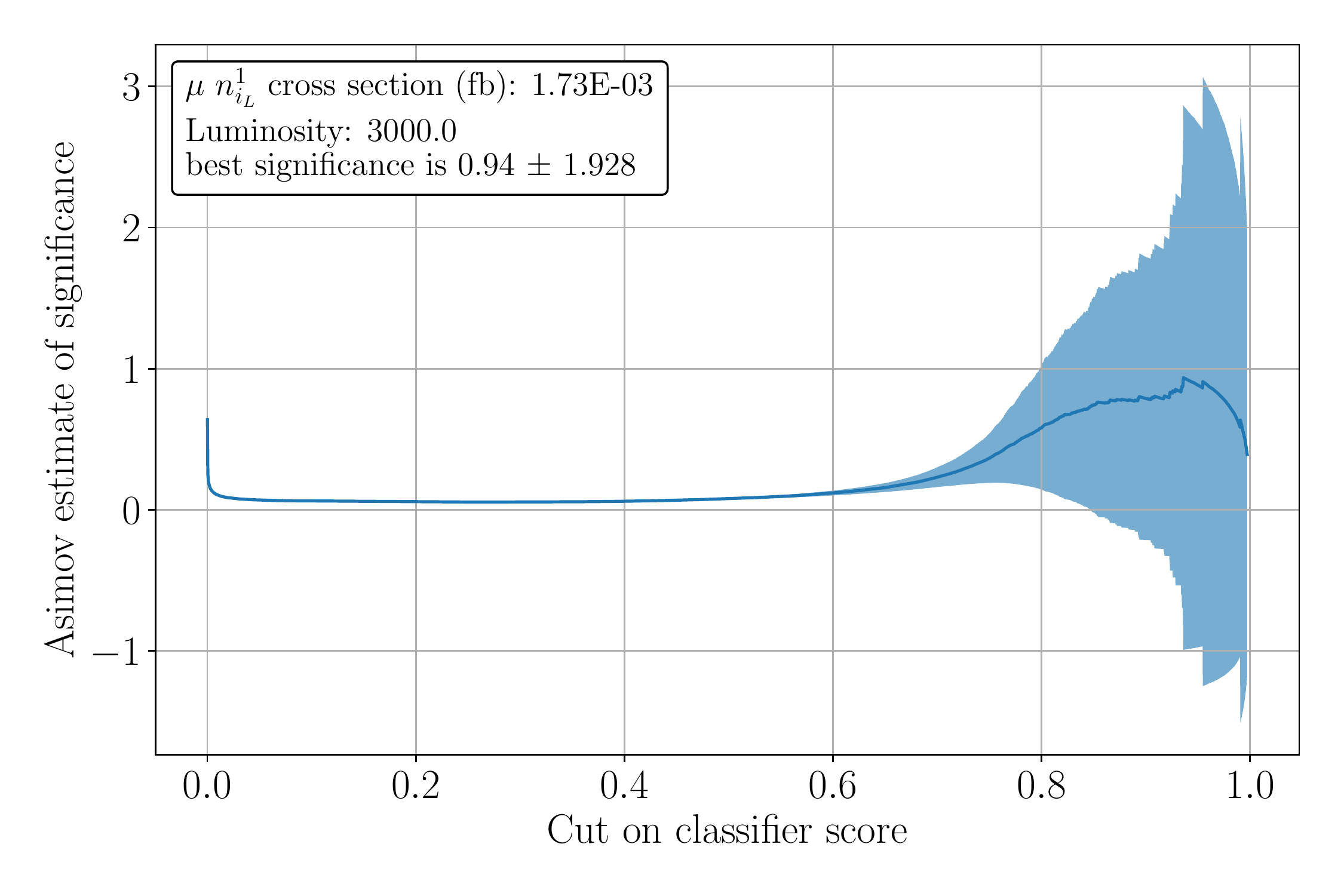}%
\includegraphics[clip,scale=0.37]{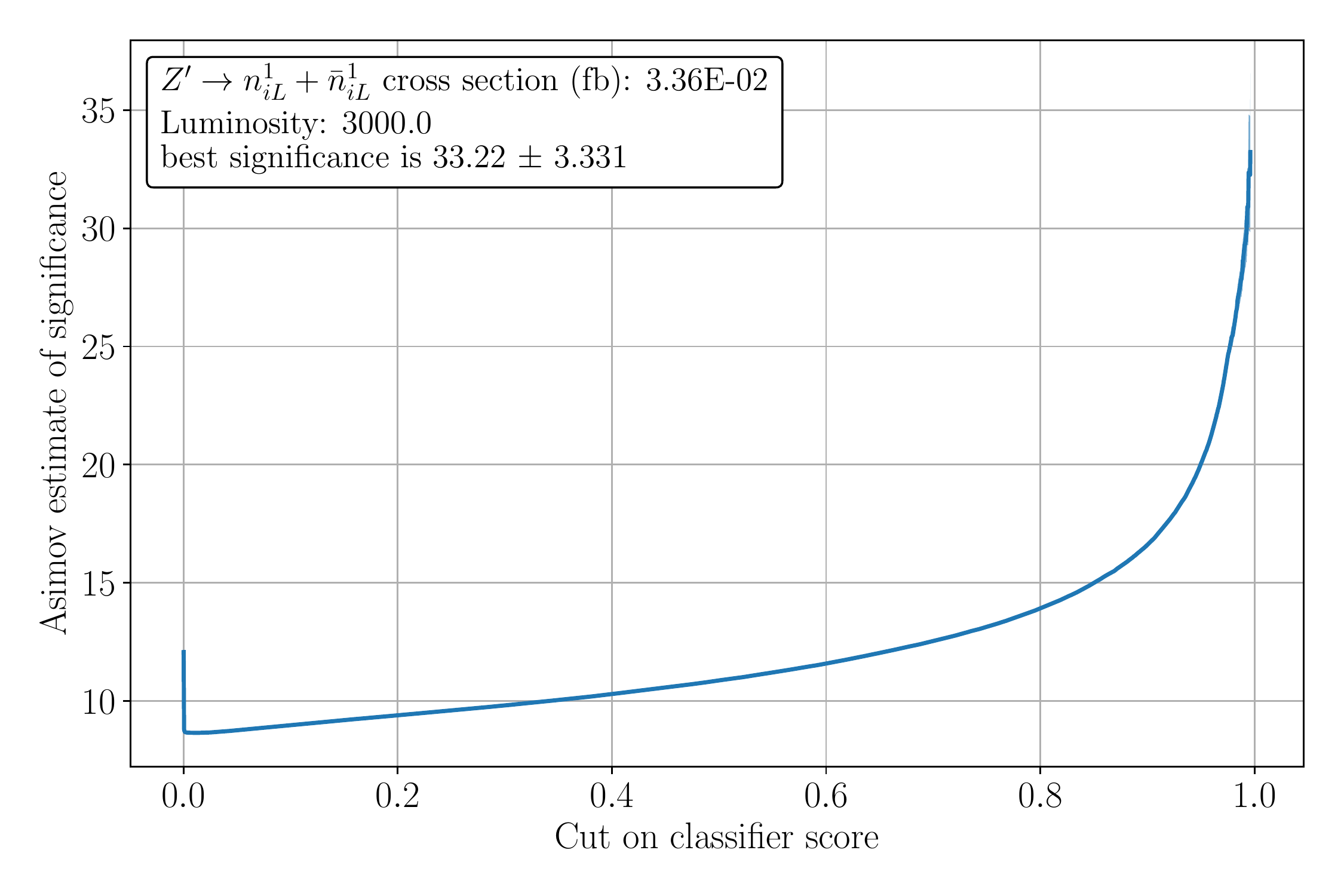}%
}

\caption{Asimov significance versus NN clasifier score for $W \rightarrow \mu \ n^1_{i_L}$ (left panel) and $n^1_{3_L}\bar{n}^1_{3_L}(n^1_{4_L}\bar{n}^1_{4_L})$ (right panel) channels for 3000 fb$^{-1}$ with 1$\%$ systematic error. The blue band represents the systematic uncertainties for the background.}
\label{Asimov_sig}
\end{center}
\end{figure}

We are interested in obtaining not only the acceptance and rejection factors, but mainly the statistical significance of the signal. To do so we can use the predictions made by our NN to estimate the number of events expected and from the number of events for each of the analysed channels get the estimate Asimov significance, which depends on the integrated luminosity and systematic uncertainties which are often disregarded in machine learning studies. The Asimov estimate of significance~\cite{Cowan:2010js}, a well-established approach to evaluate likelihood-based tests of new physics taking into account the systematic uncertainty on the background normalization, can then be used for a more careful estimate of the signal significance at the training and testing phases of construction of the classifier. The formula of the Asimov signal significance is given by
\begin{equation} \label{eq:asimov}
Z_{A} =\left[2\left((s+b)\ln\left[\frac{(s+b)(b+\bsvar)}{b^2+(s+b)\bsvar}\right]-\frac{b^2}{\bsvar}\ln\left[1+\frac{\bsvar s}{b(b+\bsvar)}\right]\right)\right]^{1/2},
\end{equation}
where, for a given integrated luminosity, $s$ is the number of signal events, $b$ is the number of background events, and the uncertainty associated with the number of background  events is given by $\sigma_b$. In Fig. \ref{Asimov_sig} we plot the estimate Asimov significance dependency over the classification score assigned by the NN. 
%%%%

Despite the relative higher cross-section for the process $pp\rightarrow W \rightarrow \mu n^{1}_{iL}$ and the 99\% accuracy achieved by the NN, the overwhelm irreducible background we have for this channel dominates the uncertainties for the Asimov significance. This imposes a bigger challenge to one who intend to probe such particle using this channel alone. Meanwhile, the process $pp\rightarrow Z' \rightarrow  n^{1}_{iL} \bar{n}^{1}_{iL}$ offers a new window to probe not only the $n^{1}_{iL}$ but the aforementioned $Z
^{\prime}$ boson. The smaller backgrounds cross-section and the 100\% accuracy achieved by the NN allow us to safely probe this channel and estimate higher significance using current LHC luminosity. Combining all these factors if the $Z^{\prime}$ is not discovery in this channel, we can exclude this model with a $Z^{\prime}$ mass below 4 TeV using current LHC luminosity. However, from FIG. \eqref{zp_xsec} we still have a wide range of mass to explore and use the analysis we developed so far as main guideline to constrain the parameters of the 331RHN.

\begin{table}[t!]
\centering
\begin{tabular}{c|c|c|c|c|c}
\hline
Systematics & 100 fb$^{-1}$ & 300 fb$^{-1}$ & 
1000 fb$^{-1}$ & 3000 fb$^{-1}$ & ATLAS+CMS combined(3 ab$^{-1})$\\ \hline\hline
1\% & \makecell{$0.17 \pm 0.352$ \\ $6.07 \pm 0.608$} & \makecell{$0.30 \pm 0.610$ \\ $10.51 \pm 1.053$} & \makecell{$0.54 \pm 1.113$ \\ $19.18 \pm 1.923$} & \makecell{$0.94 \pm 1.928$ \\ $33.22 \pm 3.331$} & \makecell{$1.95 \pm 4.0$ $(W\rightarrow \mu n^{1}_{iL})$\\ $68.97 \pm 6.915$ $(Z'\rightarrow n^{1}_{iL} \bar{n}^{1}_{iL})$} \\ \hline
5\% & \makecell{$0.17 \pm 0.352$ \\ $6.07\pm 0.608$} & \makecell{$0.30 \pm 0.610$ \\ $10.50 \pm 1.054$} & \makecell{$0.54 \pm 1.113$ \\ $19.17 \pm 1.925$} & \makecell{$0.94 \pm 1.928$ \\ $33.14 \pm 3.340$} & \makecell{$1.95 \pm 4.0$ $(W\rightarrow \mu n^{1}_{iL})$ \\ $68.88 \pm 6.922$ $(Z'\rightarrow n^{1}_{iL} \bar{n}^{1}_{iL})$} \\ \hline
10\% & \makecell{$0.17 \pm 0.352$ \\ $6.06 \pm 0.608$} & \makecell{$0.30 \pm 0.610$ \\ $10.50 \pm 1.054$} & \makecell{$0.54 \pm 1.114$ \\ $19.11 \pm 1.930$} & \makecell{$0.94 \pm 1.928$ \\ $32.90 \pm 3.364$} & \makecell{$1.95 \pm 4.0$ $(W\rightarrow \mu n^{1}_{iL})$ \\ $68.57 \pm 6.956$ $(Z'\rightarrow n^{1}_{iL} \bar{n}^{1}_{iL})$} \\ \hline
\end{tabular}
\caption[]{Projected Asimov significance of Eq.~\eqref{eq:asimov} for integrated luminosities of 100, 300, 1000 and 3000 fb$^{-1}$ at the 14 TeV LHC for the given systematic uncertainty. In the last column we show the naive combination of both LHC experiments for an integrated luminosity of 3000 fb$^{-1}$.}
\label{tab:sys_vs_lumi}
\end{table}

\begin{figure}[h!]
\begin{center}

\subfloat{%
\includegraphics[clip,scale=0.37]{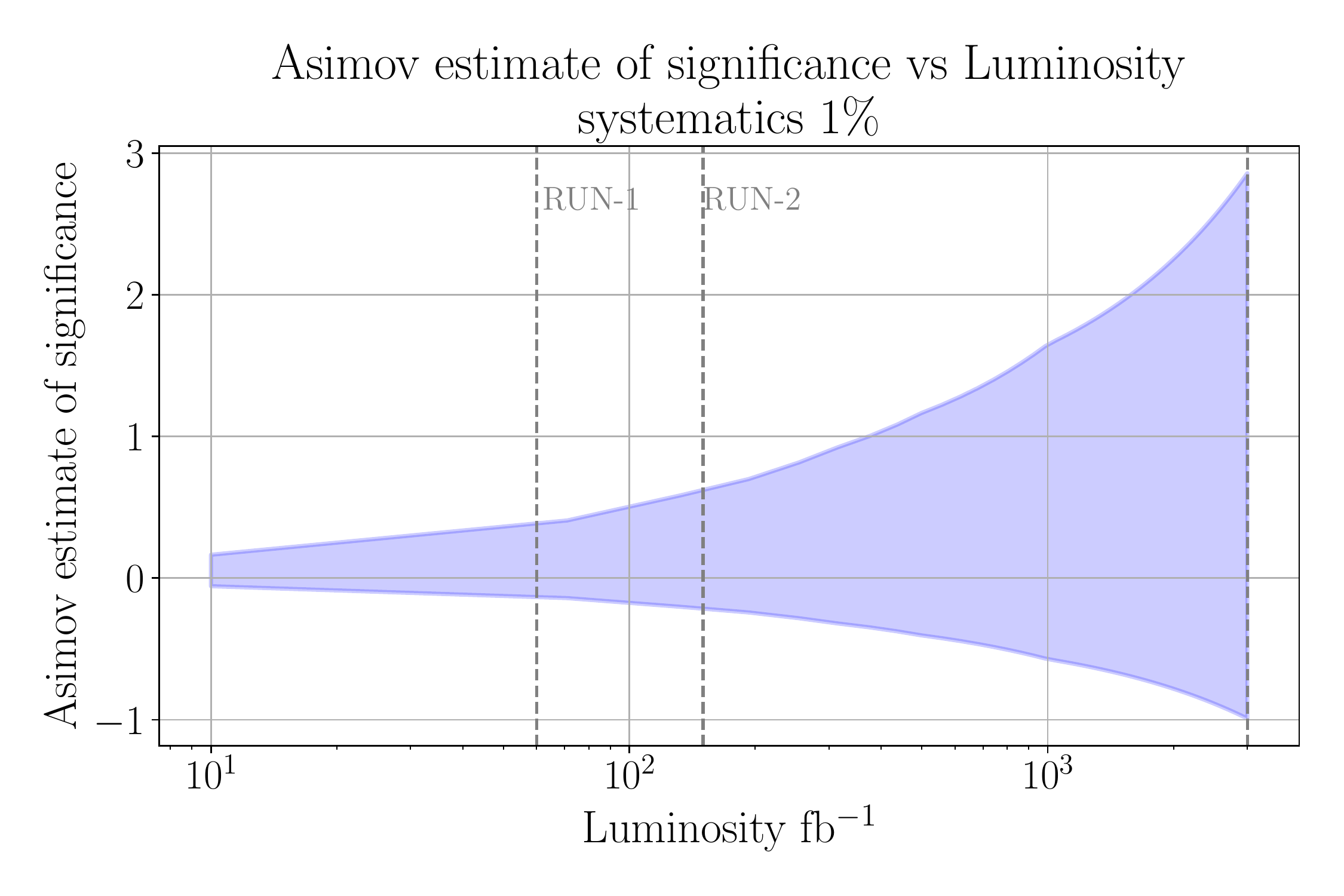}%
\includegraphics[clip,scale=0.37]{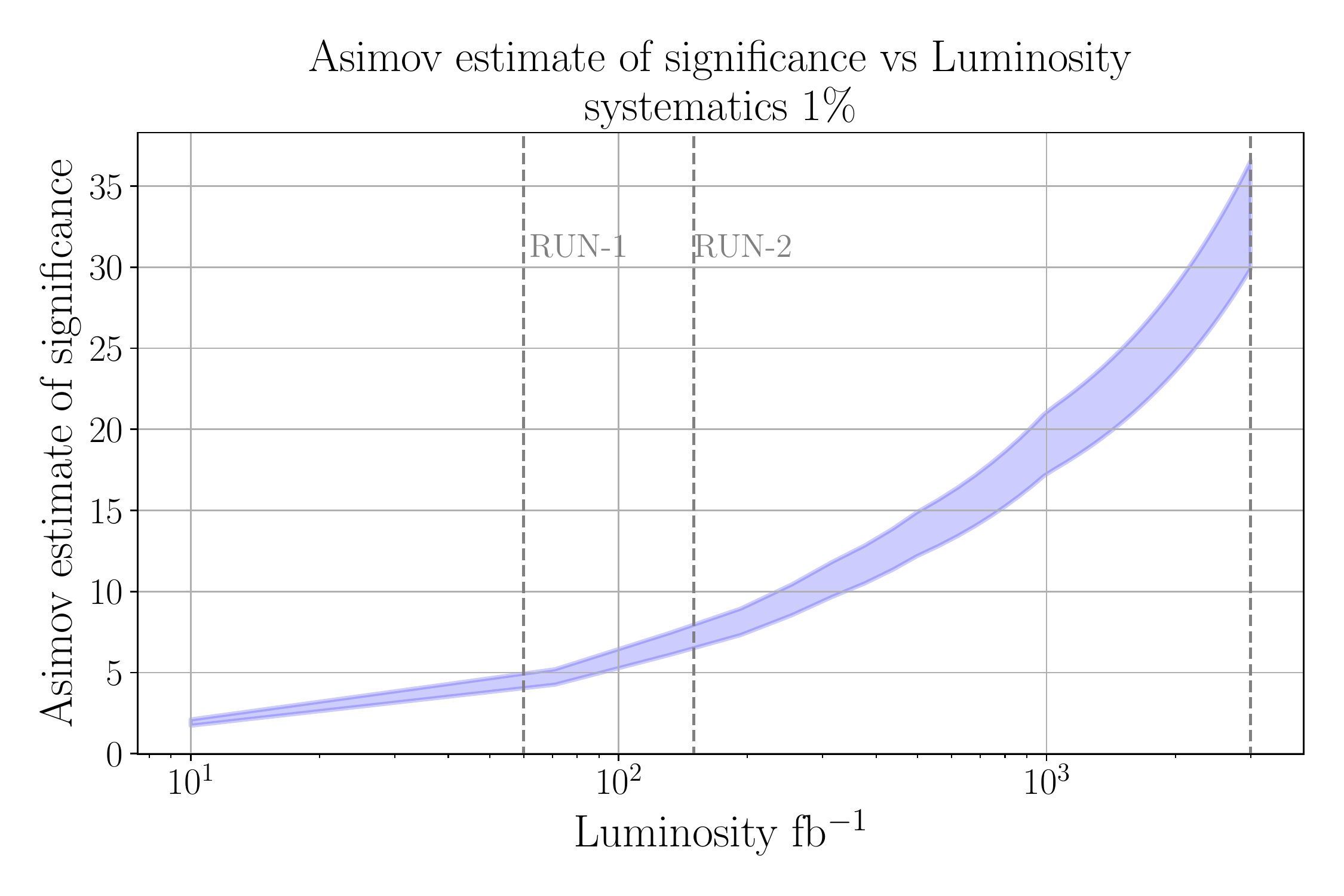}%
}

%\subfloat[ for $n^1_{3_L}\bar{n}^1_{3_L}(n^1_{4_L}\bar{n}^1_{4_L})$]{%
%\includegraphics[clip,scale=0.37]{figs/proj_sig_channel_Zp.pdf}%
%}

\caption{Luminosity (fb$^{-1}$) versus Asimov significance for $\mu \ n^1_{3_L}(\mu \ n^1_{4_L})$ (left panel) and $n^1_{3_L}\bar{n}^1_{3_L}(n^1_{4_L}\bar{n}^1_{4_L})$ (right panel) channels with 1$\%$ of background systematic error. The bands correspond to 2$\sigma$ confidence level. The dashed lines show the luminosity milestones of 60 fb$^{-1}$ (RUN 1), 150 fb$^{-1}$ (RUN 2) and 3000 fb$^{-1}$.}
\label{proj_sig}
\end{center}
\end{figure}

We can project the Asimov significance for a range of luminosity values. In FIG. \eqref{proj_sig} we have the projected significance with 1$\%$ systematic error versus the expected luminosity. The bands correspond to the projected systematic uncertainties. Due to the systematic dominance over the $W\rightarrow \mu n^{1}_{iL}$ channel, we can only achieve 3$\sigma$ significance at 3000fb$^{-1}$; yet, the projected significance for the $Z^{\prime}\rightarrow n^{1}_{iL} \bar{n}^{1}_{iL}$ shows a better perspective with 10.5 $\sigma$ of significance using the RUN-2 luminosity and around 33 $\sigma$ at 3000fb$^{-1}$ showing the sensitivity power not only of the analysis we developed, but the channel $Z^{\prime}\rightarrow n^{1}_{iL} \bar{n}^{1}_{iL}$ as well.

%%%%
%%%%%%%%%%%%%%%%%%%%%
\section{Conclusions}
In this work we revisited, in details, the implementation of the inverse seesaw mechanism into the 3-3-1 model with right-handed neutrinos and, then, probed their signatures, in the form of heavy neutrinos, at the LHC by means of deep learning techniques. The spectrum of mass for these new neutrinos may vary from some hundreds of GeVs up to TeV scale.  Our analysis considered the production of such neutrinos by means of the processes  $p p \rightarrow W^{\pm} \rightarrow \mu^{\pm}n^1_{(3,4)_L}\rightarrow  \mu^{\pm}\mu^{\mp}e^{\pm}\nu_{e}$  and $pp \rightarrow Z^{\prime}\rightarrow  n^1_{(3,4)_L} n^1_{(3,4)_L} \rightarrow \mu^+ \mu^- e^+ e^- \nu_e \bar \nu_e$. We applied deep learning techniques in conjunction with evolutionary algorithms in our analysis and  concluded that the second process is much more efficient than the first one. As main result we have that the second process allows we probe not only the signal of the ISS mechanism, but also the model in question, i.e., the 331RHN. According to our analysis if the $Z^{\prime}$ is not discovery in this channel, we can exclude within 6 $\sigma$ at 95\% of confidence level this model with a $Z^{\prime}$ mass below 4 TeV using current LHC luminosity.

\clearpage
\acknowledgments

D. Cogollo is partly supported by the Brazilian National Council for Scientific and
Technological Development (CNPq), under grants 436692/2018-0. Y. M. Oviedo-Torres acknowledges the financial support from CAPES
under grants 88887.485509/2020-00. C. Pires is partly supported by the Brazilian National Council for Scientific and
Technological Development (CNPq), under grants No. 304423/2017-3. F. F. Freitas is  supported  by the project \textit{From Higgs Phenomenology to the Unification of Fundamental Interactions} PTDC/FIS-PAR/31000/2017 grant BPD-32 (19661/2019), and P. Vasconcelos was partly supported by the Brazilian National Council for Scientific and Tech-nological Development (CNPq).

%%%%%%%%%%%%%%%%%%%%%%%%%%%%%%%%%%%%%%%%%%%%%%

%%%
\end{document}